\documentclass[12pt,preprint]{aastex}
\def\degpoint{\ifmmode ^{\rm{o}}\!. \else $^{\rm{o}}\!.$\fi}

\newcommand{\ms}{\mbox{m\,s$^{-1}$}}

\newcommand{\Mjup}{\mbox{M$_{\rm Jup}$}}

\newcommand{\ltsimeq}{\raisebox{-0.6ex}{$\,\stackrel
         {\raisebox{-.2ex}{$\textstyle <$}}{\sim}\,$}}
\newcommand{\gtsimeq}{\raisebox{-0.6ex}{$\,\stackrel
         {\raisebox{-.2ex}{$\textstyle >$}}{\sim}\,$}}

\begin{document}

\title{The Anglo-Australian Planet Search XXIV: The Frequency of Jupiter 
Analogs }

\author{Robert A.~Wittenmyer\altaffilmark{1,2,3}, R.P. 
Butler\altaffilmark{4}, C.G.~Tinney\altaffilmark{1,2}, Jonathan 
Horner\altaffilmark{3,2}, B.D.~Carter\altaffilmark{3}, D.J. 
Wright\altaffilmark{1,2}, H.R.A.~Jones\altaffilmark{5}, 
J.~Bailey\altaffilmark{1,2}, Simon J.~O'Toole\altaffilmark{6} }

\altaffiltext{1}{School of Physics, University of New South Wales, 
Sydney 2052, Australia}
\altaffiltext{2}{Australian Centre for Astrobiology, University of New 
South Wales, Sydney 2052, Australia}
\altaffiltext{3}{Computational Engineering and Science Research Centre, 
University of Southern Queensland, Toowoomba, Queensland 4350, 
Australia}
\altaffiltext{4}{Department of Terrestrial Magnetism, Carnegie 
Institution of Washington, 5241 Broad Branch Road, NW, Washington, DC 
20015-1305, USA}
\altaffiltext{5}{Centre for Astrophysics Research, University of 
Hertfordshire, College Lane, Hatfield, Herts AL10 9AB, UK}
\altaffiltext{6}{Australian Astronomical Observatory, PO Box 915,
North Ryde, NSW 1670, Australia}

\email{
rob@unsw.edu.au}

\shorttitle{AAPS Jupiter Analogs }
\shortauthors{Wittenmyer et al.}

\begin{abstract}

\noindent We present updated simulations of the detectability of Jupiter 
analogs by the 17-year Anglo-Australian Planet Search.  The occurrence 
rate of Jupiter-like planets that have remained near their formation 
locations beyond the ice line is a critical datum necessary to constrain 
the details of planet formation.  It is also vital in our quest to fully 
understand how common (or rare) planetary systems like our own are in 
the Galaxy.  From a sample of 202 solar-type stars, and correcting for 
imperfect detectability on a star-by-star basis, we derive a frequency 
of $6.2^{+2.8}_{-1.6}$\% for giant planets in orbits from 3-7\,AU.  When 
a consistent definition of ``Jupiter analog'' is used, our results are 
in agreement with those from other legacy radial velocity surveys.

\end{abstract}


\section{Introduction}

Much attention has been brought to bear in recent years on the 
occurrence rate of Earth-like planets \citep[e.g.][]{howard12, etaearth, 
kop13}.  This is due in large part to the flood of data from the 
\textit{Kepler} spacecraft mission, which has provided evidence that 
small planets are exceedingly common \citep[e.g.][]{fressin13, 
dressing15, burke15}.  These findings are a critical step towards 
answering the fundamental question ``how common are planetary systems 
like our own Solar system?'' But to fully understand the degree to which 
our Solar system is unusual, we must also consider the other planets 
therein.  In other words, how common are planetary systems that feature 
distant giant planets such as our own gas and ice giants (Jupiter, 
Saturn, Uranus and Neptune).  The other half of the problem, 
then, requires understanding the frequency and properties of planets 
like our own Jupiter.


One can argue that Jupiter, as the most massive and dynamically dominant 
body, is a key component that makes our Solar system what it is today.  
Jupiter, as the most massive and dynamically dominant body, has played a 
pivotal role in shaping our Solar system into what we see today.  That 
influence can be seen in many ways when one examines the modern day 
Solar system. The Asteroid belt, interior to Jupiter's orbit, has been 
sculpted over the past four and a half billion years to display 
intricate fine structure.  The bulk of that structure is the direct 
result of perturbations from Jupiter, and, to a lesser extent, Saturn.  
Jupiter also acts to control the flux of small bodies to the inner Solar 
system, acting to perturb asteroids and comets onto Earth-crossing 
orbits \citep[e.g.][]{laakso06, hj08, hj12}.  Jupiter also hosts a large 
population of Trojan asteroids \citep{forn07, h12, v15} and irregular 
satellites, both of which are thought to have been captured during the 
giant planet's migration \citep[e.g.][]{sheppard03, morbi05, jewitt07, 
l10}.  The planet has even been put forward as having played a key role 
in the volatilisation of the terrestrial planets, driving the injection 
of a late veneer of volatile material to the inner Solar system 
\citep{owen95, hj10}, and helps to drive periodic climate change on the 
Earth, in the form of the Milankovitch cycles \citep{hays76, horner15}.

Given Jupiter's importance in the creation of the Solar system as we 
observe it - the only planetary system known to host life - it is 
clearly important to constrain the frequency of Jupiter analogs when 
studying the question of our Solar system's ubiquity.  In estimating the 
frequency of Jupiter-like planets in Jupiter-like orbits, we must first 
define a ``Jupiter analog.'' A reasonable and physically-motivated 
definition is as follows: a gas-giant planet that plays a similar 
dynamical role to our own Jupiter, and that lies beyond the ice line.  

The first criterion sets a lower bound on the planetary mass -- a Saturn 
mass (0.3\,\Mjup) is a reasonable limit, though in practice the 
sensitivity of Doppler radial velocity surveys at present obviates the 
need to set an explicit lower bound here.  A Saturn-mass planet in a 
10-year orbit about a Solar-type star has a velocity amplitude of 
4\,\ms, a signal currently at the edge of detectability for long-running 
``legacy'' radial velocity surveys which have a typical velocity 
precision of 2-3\,\ms\ per epoch.  A more physically-motivated lower 
mass boundary may be half of a Saturn mass ($\sim$0.15\,\Mjup), 
corresponding to the overturn in the frequency of impacts that would be 
experienced by an Earth-like planet from a regime increasing with 
Jupiter-mass to one decreasing \citep{hj08, hj09}.  We set an 
upper mass limit of 13\Mjup, consistent with the accepted boundary 
between planets and brown dwarfs.

The second criterion ensures that such a planet has not migrated 
significantly beyond its formation location, leaving dynamical room for 
interior potentially rocky, habitable planets.  This sets an inner limit 
of $\sim$3\,AU, which has been used by previous studies of Jupiter 
analogs \citep{jupiters, rowan15}.  Giant planets that stay beyond this 
point should not prevent the accretion of telluric worlds.  Finally, we 
require such a planet to have a low eccentricity ($e\ltsimeq$0.3), 
indicating that the system has had a relatively benign dynamical 
history, preserving any interior rocky planets.  Table~\ref{jupiters} 
gives a list of Jupiter analogs in the AAPS according to this 
definition.

The occurrence rate of Jupiter analogs has been estimated from radial 
velocity surveys \citep[e.g.][]{cumming08, jupiters, rowan15} and from 
microlensing \citep{gould10}.  The former studies have generally arrived 
at a Jupiter-analog frequency of $\sim$3-4\% (agreeing with each other 
within uncertainties) whilst the latter arrived at a Solar system analog 
frequency of $\sim$17\% based on one detection of a 
Jupiter/Saturn analog pair.  As the temporal duration of radial 
velocity survey baselines has increased, we are beginning to be able to 
access orbits with semi-major axes of $a\gtsimeq$6-8\,AU.  At the same 
time, advanced direct-imaging instruments such as the Gemini Planet 
Imager and VLT/SPHERE are now able to probe inward of $\sim$10\,AU 
\citep{zurlo15, vigan15}, the coming decade will see great advances in 
our understanding of not only the frequency, but also the properties of 
Jupiter-like planets in Jupiter-like orbits.

In this paper, we expand on our work in \citet{jupiters}, adding a 
further 5 years of observational data from the Anglo-Australian Planet 
Search (AAPS), which has now been in continuous operation for 17 years. 
This allows us to deliver a refined estimate of the occurrence rate of 
Jupiter analogs in our sample.  Section 2 describes the input data 
properties and numerical methods.  Results are given in Section 3, and 
in Section 4 we give our conclusions.

\section{Observations and Computational Methods}

The Anglo-Australian Planet Search (AAPS) has been in operation since 
1998 January, and monitored about 250 stars for the first 14 years.  
Since 2013, the AAPS has refined its target list to the $\sim$120 stars 
most amenable to the detection of Jupiter analogs.  This is in response 
to simulation work in \citet{jupiters} and \citet{witt13} which 
identified the most favourable and most active stars.  Increasingly 
limited telescope time also required the AAPS to drop targets which had 
too few observations to ``catch up.'' The AAPS has achieved a long-term 
radial-velocity precision of 3\,\ms\ or better since its inception, 
which enables the detection of long-period giant planets.  Indeed, the 
detection of such planets is a strength of the AAPS; of the 40 planets 
discovered by the AAPS, 16 (40\%) have orbital periods longer than 1000 
days.

\begin{figure}
\plotone{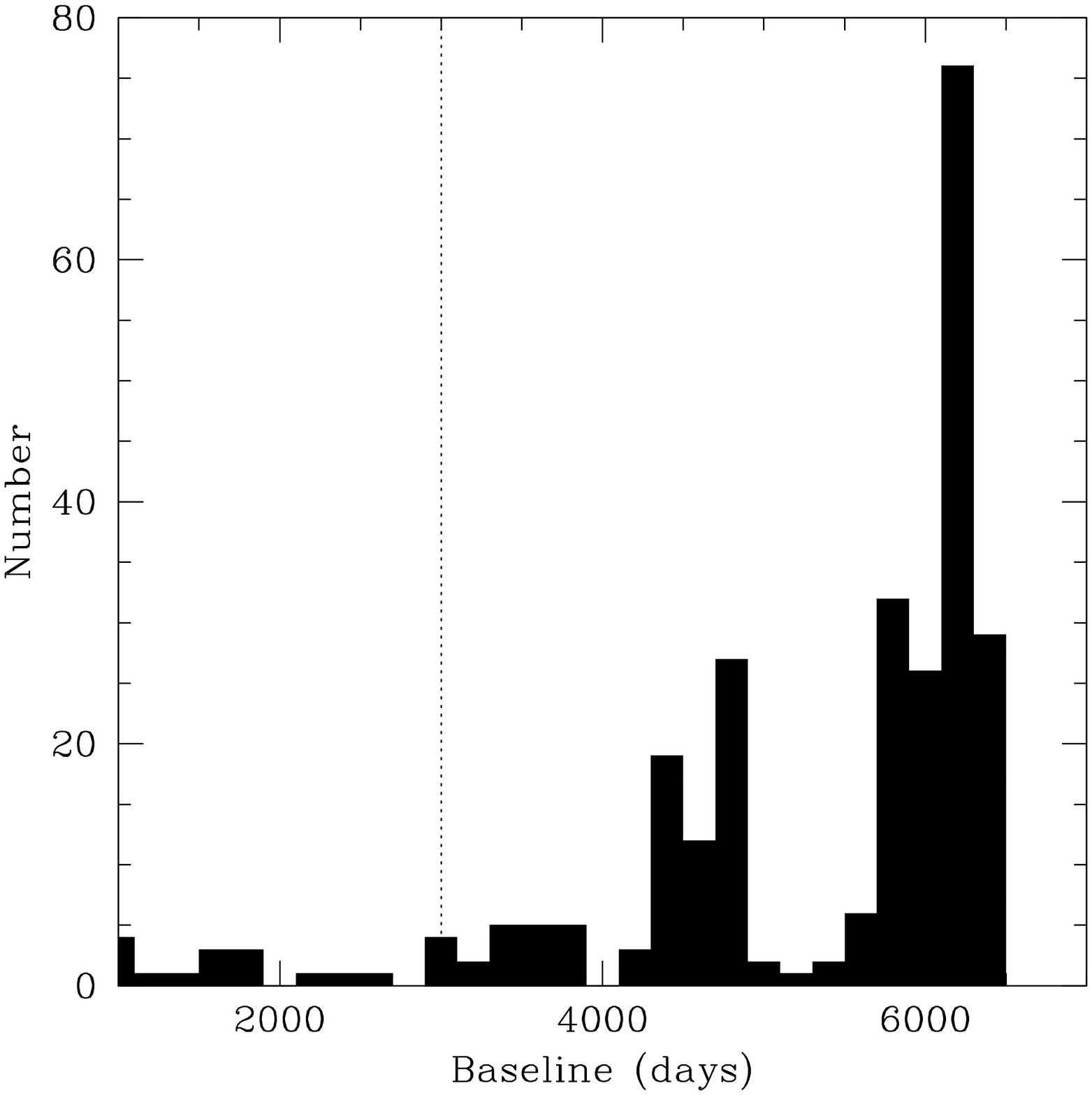}
\caption{Histogram of observational baselines for 271 stars from the AAPS.  The majority of our targets have been observed for 
at least 6000 days (i.e.\,$>16.4$ years).  The vertical dashed line shows our requirement of at least 3000 days of data for this 
analysis. }
\label{baselines}
\end{figure}

To determine the underlying frequency of Jupiter analogs (defined above 
as planets with $a>3$\,AU, m sin $i>$0.3\,\Mjup, and $e\ltsimeq0.3$) we 
apply the selection criteria used in \citet{jupiters}.  That is, we only 
consider those AAPS targets which have more than 8 years of data and at 
least $N=30$ observations.  The first criterion ensures that there is 
sufficient observational baseline to detect a Jupiter analog through its 
complete orbit, and the second criterion improves the reliability of the 
false-alarm probability (FAP) estimation used in our detection-limit 
technique.  Figure~\ref{baselines} shows a histogram of the time 
baselines for all 271 AAPS stars; nearly all have $T_{obs}>$3000 days 
and most have $T_{obs}>$6000 days.  After applying the selection 
criteria above, we have 202 AAPS stars which will constitute the 
Jupiter-analog sample hereafter.  Table~\ref{rvdata1} summarises the 
data characteristics for these 202 stars.  For those stars with 
long-term trends, a linear or quadratic fit was removed from the data 
before subjecting them to our detection-limit procedure.  For stars 
known to host a substellar companion, we fit for and removed that orbit 
and then performed the detection-limit computations on the residuals.

\begin{figure}
\plottwo{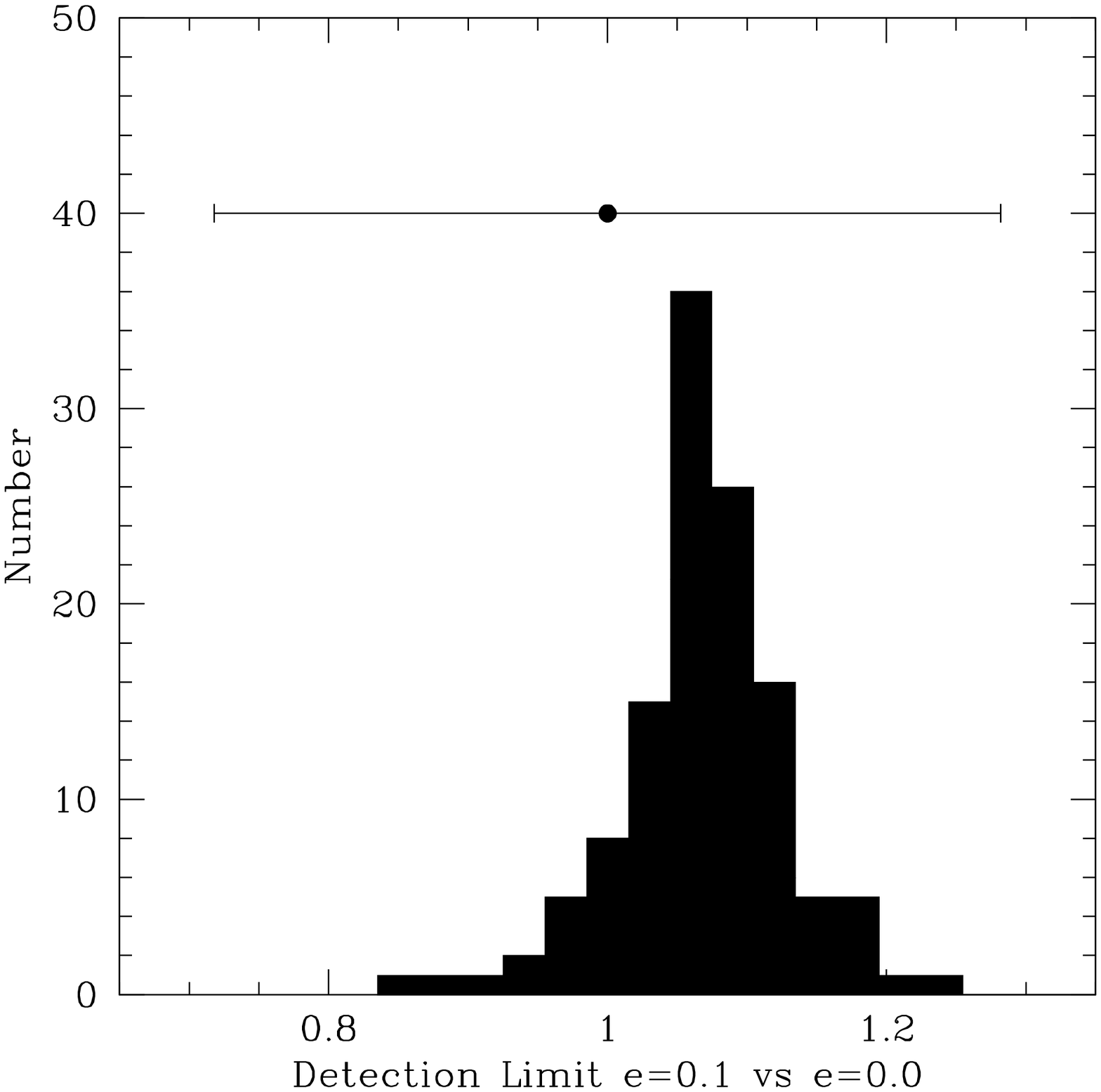}{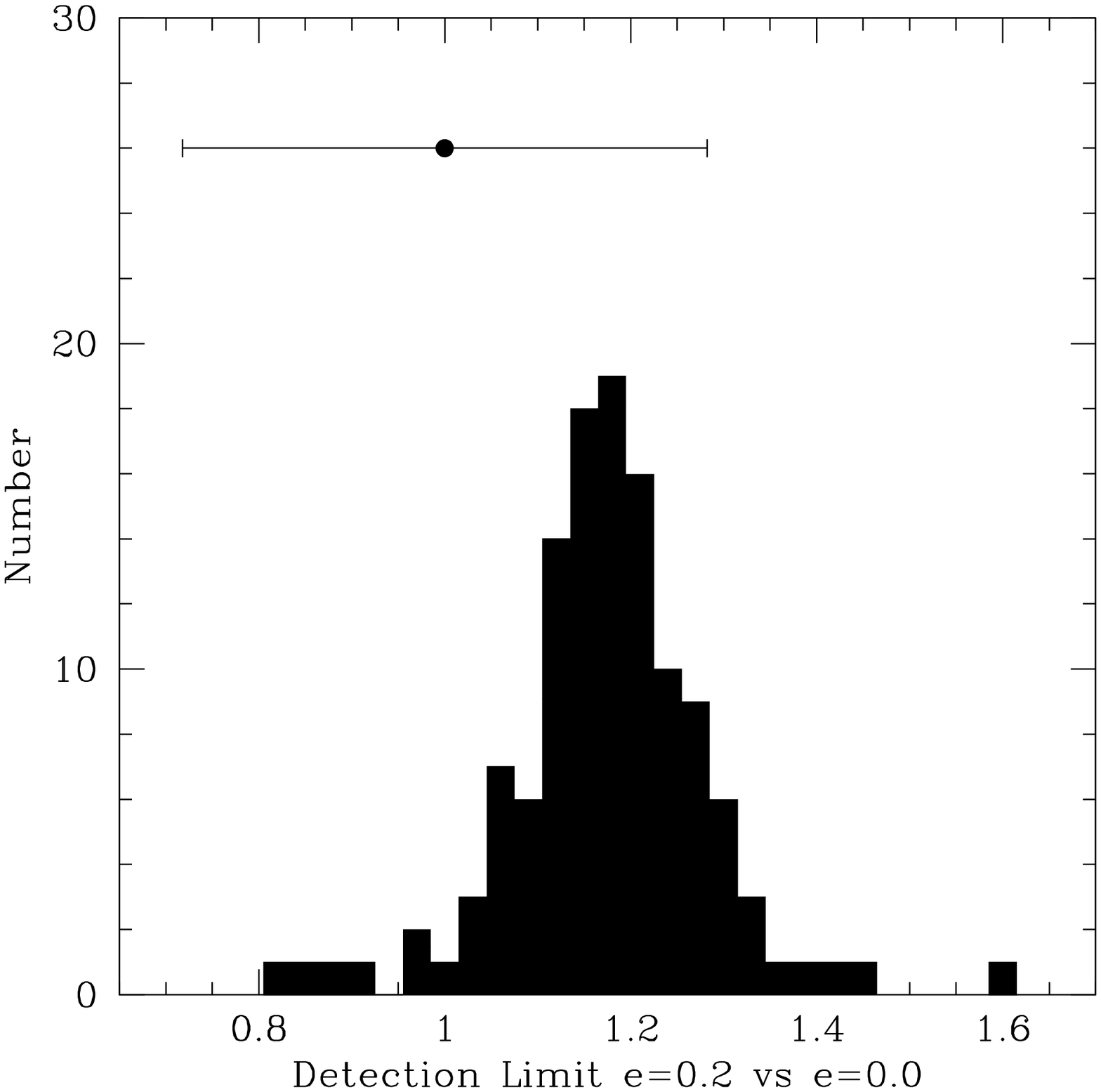}
\caption{Distribution of mean velocity amplitude detectable at 99\% 
confidence ($\bar{K}$) from Table 2 of \citet{jupiters}.  Left panel: 
$e=0.1$; right panel: $e=0.2$.  The error bar represents the fractional 
uncertainty in $\bar{K}$ as averaged over the trial periods from 
1000-6000 days.  For the small eccentricities considered for Jupiter 
analogs in this work ($e<0.3$), the effect of eccentric orbits is 
within the uncertainties of $\bar{K}$.  }
\label{ecc1}
\end{figure}

We derived detection limits using the same technique as in 
\citet{jupiters} and other work by our group 
\citep[e.g.][]{foreverpaper, etaearth, debris}.  In brief, the Keplerian 
orbit of an artificial planet is added to the data, then we attempt to 
recover that signal using a generalised Lomb-Scargle periodogram 
\citep{zk09}.  A planet is considered detected if it is recovered with 
FAP$<$1\% based on the FAP estimation in \citet{zk09}.  Comparisons of 
the FAP thresholds achieved by this analytic approach to those derived 
from a full bootstrap randomisation \citep{kurster97} have verified that 
the two methods give consistent results.  We considered planets with 100 
trial orbital periods between 1000-6000 days.  The detection limit has 
been shown to be only minimally sensitive to small nonzero 
eccentricities ($e<0.5$; Cumming et al.~2010).  In \citet{jupiters}, we 
also derived the detection limits at $e=0.1$ and $e=0.2$.  To illustrate 
the effect of (small) eccentricities on the radial velocity amplitude 
$K$ detectable in AAT data, we revisit the results of \citet{jupiters}.  
Figure~\ref{ecc1} shows the distribution of the mean detection limit 
$\bar{K}$ for the 123 stars in that work (results taken from their Table 
2).  For $e=0.1$, the detection limit increased by only $\sim$5\%, while 
for $e=0.2$ the limit increased by $\sim$15\%.  The typical uncertainty 
in $\bar{K}$ is comparable to these eccentricity effects, and so we 
conclude that for the low-eccentricity orbits of Jupiter analogs (as 
defined above), the circular case is sufficiently informative.  Hence, 
we consider only circular orbits in this work.

\section{Results}

\subsection{Detection Limits}

\begin{figure}
\plotone{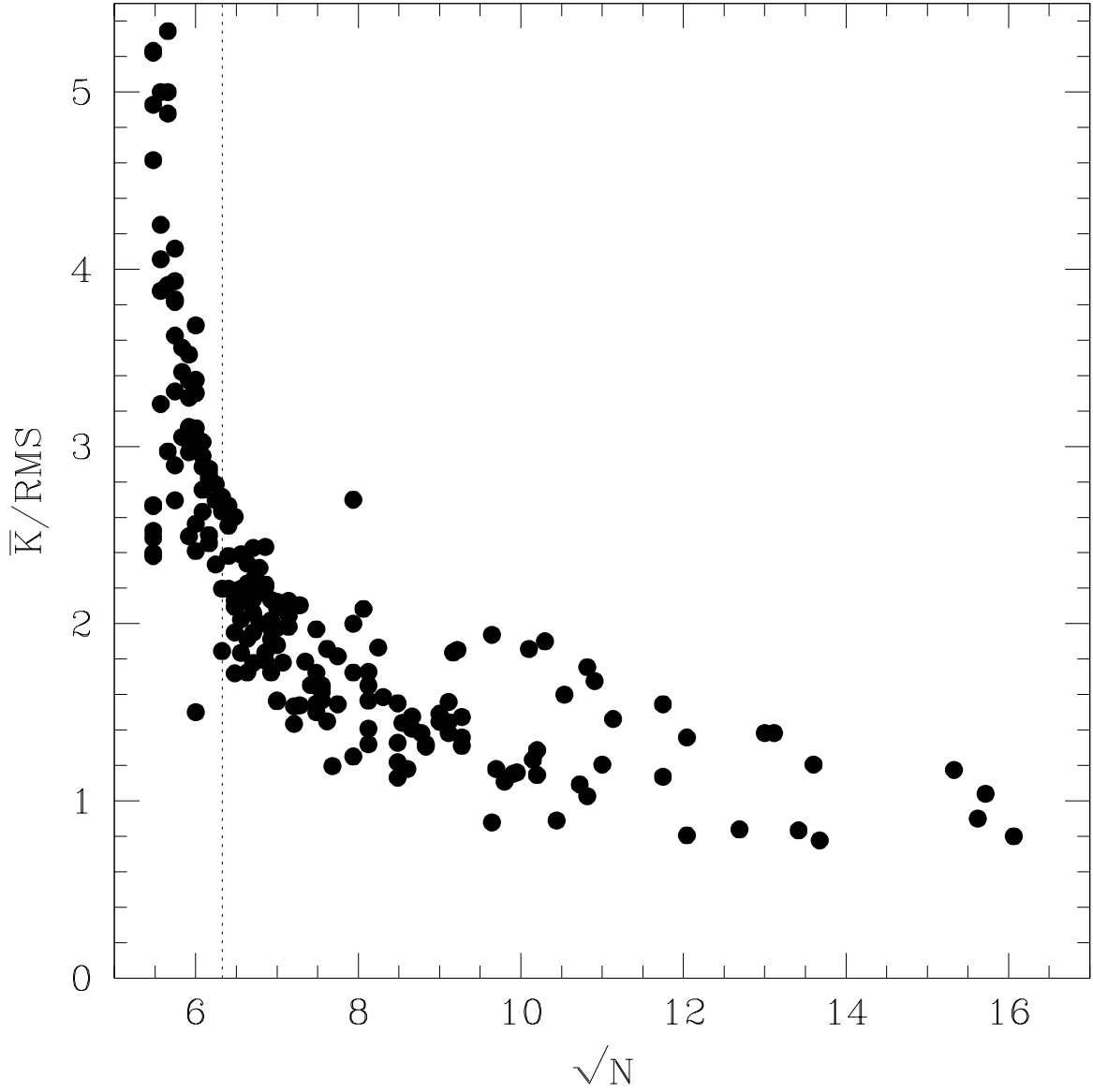}
\caption{Normalised detection limit at 99\% recovery ($\bar{K}$/RMS) 
versus $\sqrt{N}$ for our 202 stars.  The dashed line indicates $N=40$, 
a reasonable minimum number of observations required to obtain a 
detection limit that scales as the expected $\sqrt{N}$. }
\label{goingtoshit}
\end{figure}

Complete results for all 202 stars are given in Table~\ref{meankays}, 
which shows the mean velocity amplitude $\bar{K}$ detectable at six 
recovery rates: 99\%, 90\%, 70\%, 50\%, 30\%, and 10\%.  There are 
substantial differences from one star to the next.  We can normalise the 
results to each star's intrinsic RMS scatter by considering the quantity 
$\bar{K}/RMS$.  This allows us to express the dependence of our achieved 
detection limit (sensitivity) on the number of observations $N$ -- a 
figure of merit which can be useful in planning how many additional 
observations are required to obtain a robust detection (or 
non-detection) of a particular class of planet.  
Figure~\ref{goingtoshit} shows the result of this exercise.  We plot 
$\bar{K}/RMS$ versus $\sqrt{N}$, with the expectation that the detection 
limit should show a linear relationship with $\sqrt{N}$.  In previous 
work on detection limits, we have required $N>30$ based on experience 
with FAP computations which become unreliable for small samples.  It is 
evident from Figure~\ref{goingtoshit} that a better choice would be 
$N>40$; that is the region where the relation between normalised 
detection limit $\bar{K}/RMS$ and $\sqrt{N}$ displays the expected 
linear relationship.  We can then define the following relation for 
$N>40$:

\begin{equation}
\frac{\bar{K}}{RMS} = 2.28\pm 0.42 - \Big{(}0.093\pm 0.036\Big{)}\sqrt{N}, 
\end{equation}


\noindent where the symbols have their usual meaning.  If the goal is a 
detection limit equal to the RMS scatter of the velocity data, then 
approximately $N=190$ observations would be required.  As legacy radial 
velocity searches such as the AAPS, Texas, and California Planet Survey 
\citep{2jupiters, endl15, fischer14} extend their time baselines toward 
Saturn-like orbits ($P_{Saturn}=$29\,yr), then a long-term precision of 
3\,\ms\ or better is required to detect or exclude a Saturn analog 
($K_{Saturn}=3$\,\ms).


\subsection{The Frequency of Jupiter Analogs}

The primary aim of this work is to derive the frequency of Jupiter 
analogs in the AAPS sample.  In this sample of 202 stars, a total of 8 
Jupiter analogs (per the criteria in the Introduction) have been 
detected to date; their properties are enumerated in 
Table~\ref{jupiters}.  We have excluded HD\,39091b as its eccentricity 
$e=0.638\pm$0.004 is well beyond our definition of a Jupiter analog.  
Such a highly eccentric planet is likely to have resulted from severe 
dynamical interactions \citep{ford08, chat08}, rendering the HD\,39091 
system almost certainly \textit{not} Solar system-like.  In addition, 
previous studies have shown that some such moderate-high eccentricity 
planets turn out, on later investigation, to be two planets on circular 
orbits \citep{142paper, songhu}.

\begin{figure}
\plotone{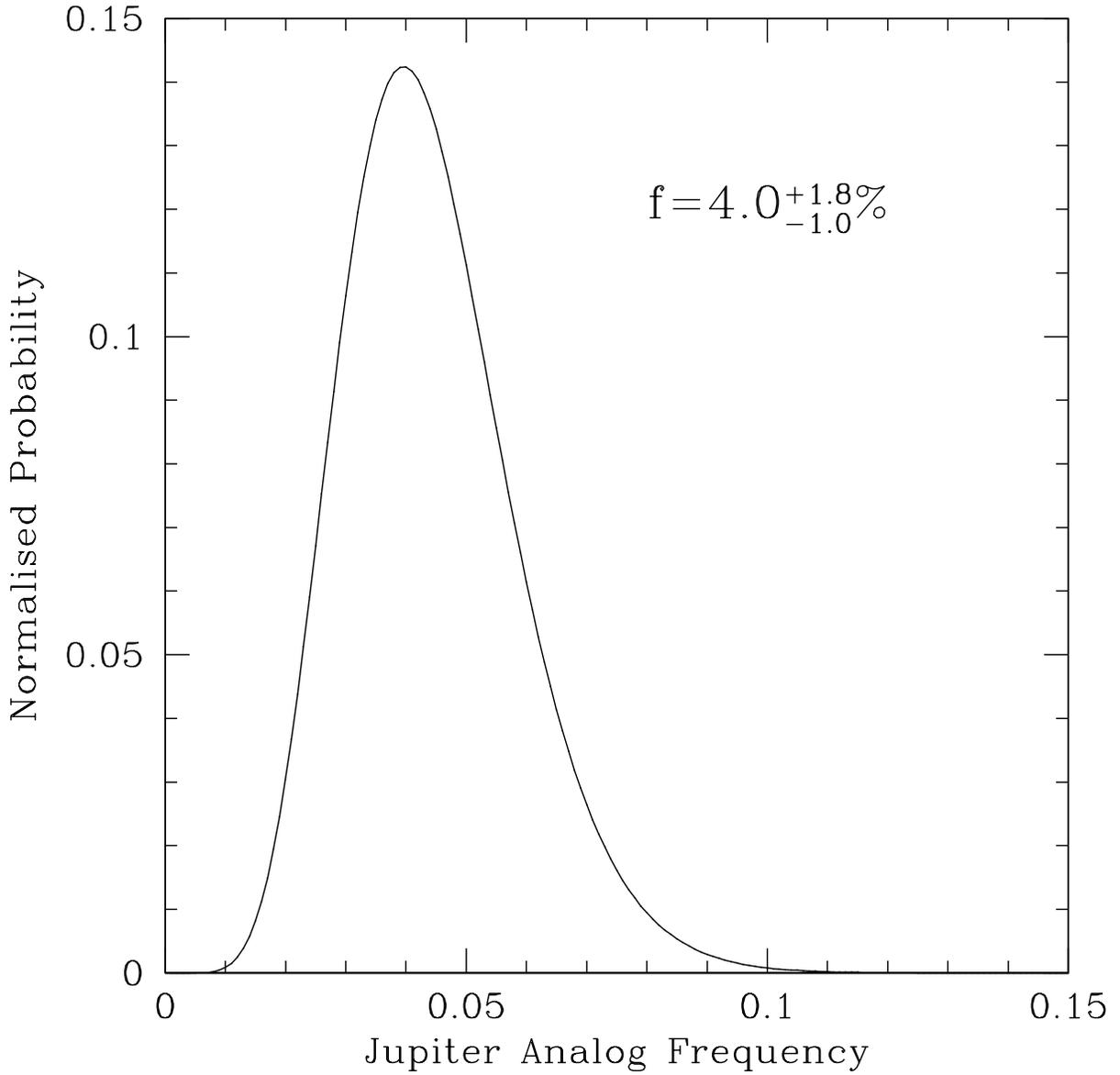}
\caption{Binomial probability density function for Jupiter analogs based 
on the 8 detections in the 202-star AAPS sample. This yields a frequency 
of $f=4.0^{+1.8}_{-1.0}\%$ (uncorrected for imperfect detectability). }
\label{probs}
\end{figure}

\begin{figure}
\plotone{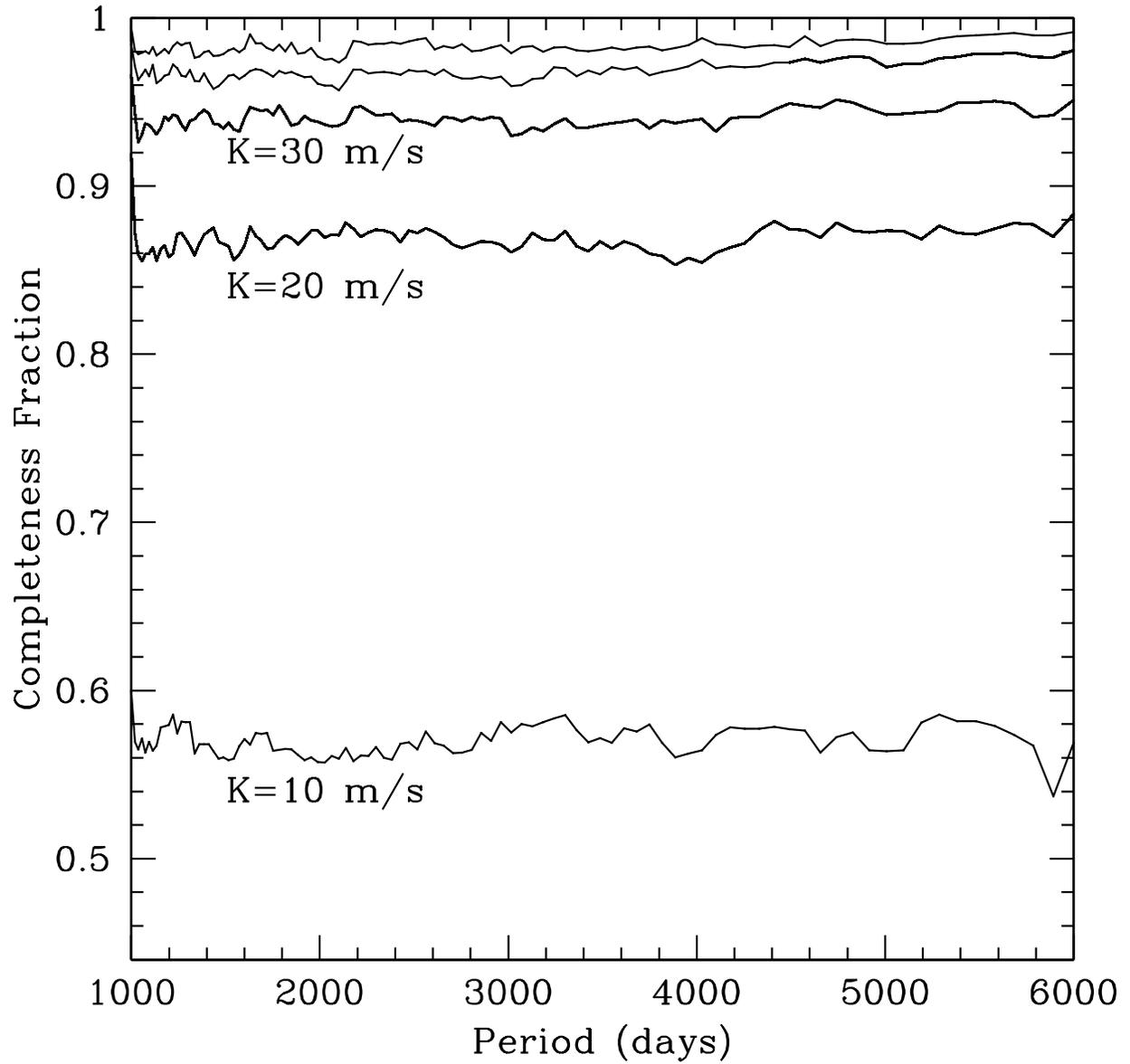}
\caption{Completeness fraction for 202 AAPS stars, as a function of 
orbital period and radial-velocity amplitude $K$. From bottom to top, 
the curves are for $K=$10, 20, 30, 40, and 50 \ms. }
\label{complete}
\end{figure}

Following previous work \citep{howard10, etaearth}, we can use binomial 
statistics to estimate the frequency of Jupiter analogs in our sample.  
We compute the binomial probability of detecting exactly $k$ planets in 
a sample of $n$ stars, with the underlying probability $p$ of hosting a 
planet.  Figure~\ref{probs} shows the probability distribution based on 
detecting 8 planets in a sample of 202 stars.  This calculation yields a 
Jupiter analog frequency of $4.0^{+1.8}_{-1.0}$\%, where the uncertainty 
is the 68.3\% confidence interval about the peak of the distribution.  
However, this calculation includes no information about the relative 
detectability of such planets.  If Jupiter analogs were perfectly 
detectable for all stars in this sample, the frequency of such planets 
would simply be $8/202=4.0\%$.  To compute the true underlying frequency 
of Jupiter analogs, we must correct the sample for incompleteness, as 
done previously by \citet{jupiters, etaearth}.  This is essentially 
asking how many planets could have been missed.  We can then adjust the 
binomial results above by multiplying the Jupiter analog frequency and 
its uncertainty by a factor $(N_{detected}+N_{missed})/N_{detected}$.  
Following \citet{jupiters}, we define the survey completeness for a 
given radial-velocity amplitude $K$ and period $P$ as:

\begin{equation}
f_{c}(P,K) = \frac{1}{N_{stars}}\sum_{i=1}^{N} f_{R}(P_i,K_i),
\end{equation}

\noindent where $f_R(P,K)$ is the recovery rate as a function of $K$ at 
period $P$, and $N$ is the total number of stars in the sample 
($N=202$).  In this way, we account for the detectabilities for each 
star individually, at each of the 100 trial periods.  We use the 
specific detection limit $K_P$ obtained for each period from the 
simulations described above, thus generating six pairs of ($K_P$, 
recovery fraction).  Then, we generate $f_R(P,K)$ for each star by 
performing a linear interpolation between the six pairs of ($K_P$, 
recovery fraction).  We can then estimate the recovery fraction 
$f_R(P,K)$ for any $K$.  Under this scheme, an extremely stable star 
would have $f_R(P,K)=1.0$, representing 100\% detectability for a given 
$(P,K)$ pair.  Conversely, a star with poor detection limits would have 
a small value of $f_R(P,K)$ -- approaching zero for an exceptionally 
``bad'' star or for small $K$.  Figure~\ref{complete} shows the survey 
completeness obtained by summing over all 202 stars, for a range of 
amplitudes $K$ from 10\,\ms\ to 50\,\ms.

The completeness fraction in Equation~(2) can be used to derive a 
completeness correction for the published detections of Jupiter analogs 
in the AAPS sample.  For each of the eight stars hosting a Jupiter 
analog (Table~\ref{jupiters}), we can compute $f_R(P,K)$ at the specific 
values of $P$ and $K$ for that known planet.  All of these planets are 
100\% detectable based on the current AAT data for those stars.  The 
frequency of Jupiter analogs based on this sample, corrected for 
completeness (detectability), is then given by

\begin{equation}
f_{Jup} = \frac{1}{N_{stars}}\sum_{i=1}^{N_{hosts}} 
\frac{1}{f_{R,i}(P_i,K_i)f_{c}(P_i,K_i)} = 6.2\%.
\end{equation}

\noindent Here, $N_{stars}=202$ total stars in the sample, $N_{hosts}=8$ 
which host a Jupiter analog, and $f_R(P_i,K_i)$ refers to the recovery 
fractions listed above.  In addition, $f_c(P_i,K_i)$ (Equation 2) is 
summed over the 194 stars which did not host a Jupiter analog, to 
account for how detectable the eight found planets would have been 
around the remaining stars in the sample.  We estimate from Equation~(3) 
that 4.55 planets were ``missed,'' giving a completeness correction of 
$(N_{detected}+N_{missed})/N_{detected} = 1.56$.  Hence, correcting the 
binomial results above yields a Jupiter analog frequency of 
$6.2^{+2.8}_{-1.6}$\%.

\section{Discussion and Conclusions}

The frequency of Jupiter analogs has been estimated by several authors 
using both radial velocity and microlensing results.  Our results are 
consistent with the literature, i.e.~that Jupiter-like planets in 
Jupiter-like orbits are relatively uncommon, occurring around less than 
10\% of stars.  For giant planets beyond 3\,AU, frequencies of 
$f\,\sim$3\% are reported by several teams \citep{cumming08, jupiters, 
rowan15}, with uncertainties of $1-3\%$.  The recent work of 
\citet{rowan15} (R15), whose techniques most closely mirror our own, 
resulted in an estimate of $f\sim\,1-4\%$ (90\% confidence interval).  
At first glance, this is in disagreement with our result of 
$6.2^{+2.8}_{-1.6}$\%.  However, we note two important differences: (1) 
R15 defined Jupiter analogs as planets with masses 0.3-3\,\Mjup\ and 
periods 5-15 years, and (2) they report a 10-90\% confidence interval 
whereas we report a 68.7\% (1$\sigma$) confidence interval about the 
peak of the posterior distribution function.  If we adopt the Jupiter 
analog definition of R15, HD\,142c and HD\,30177b no longer count, and 
we get a binomial probability of 6 detections in 202 stars as 
$f\sim\,2.0-4.7\%$ (90\% confidence interval).  Correcting for missed 
planets as in Equation (3), this range becomes $f\sim\,3.1-7.3\%$.  
Hence, by aligning our definitions and reported confidence intervals, 
our results overlap with those of R15.


Our central value for the Jupiter analog frequency remains somewhat 
higher, which can be attributed to the missed-planet correction.  While 
both this work and R15 determined survey completeness via injection and 
recovery simulations, R15 averaged detectability over phases at a given 
period, whereas our technique considered recovery as a binary function - 
that is, if one of the 30 trial phases resulted in a non-detection, the 
$K$ amplitude at that period was deemed ``not recovered'' and $K$ was 
increased until all phases resulted in a significant recovery of the 
injected signal.  The result is that our approach would give higher 
(more conservative) limits and, by Equation (2), that translates into 
lower recovery rates for a given $K$, leading to a larger missed-planet 
correction (Eq.~(3)) than that derived by R15.  This in turn leads to a 
higher Jupiter analog frequency.

\begin{figure}
\plotone{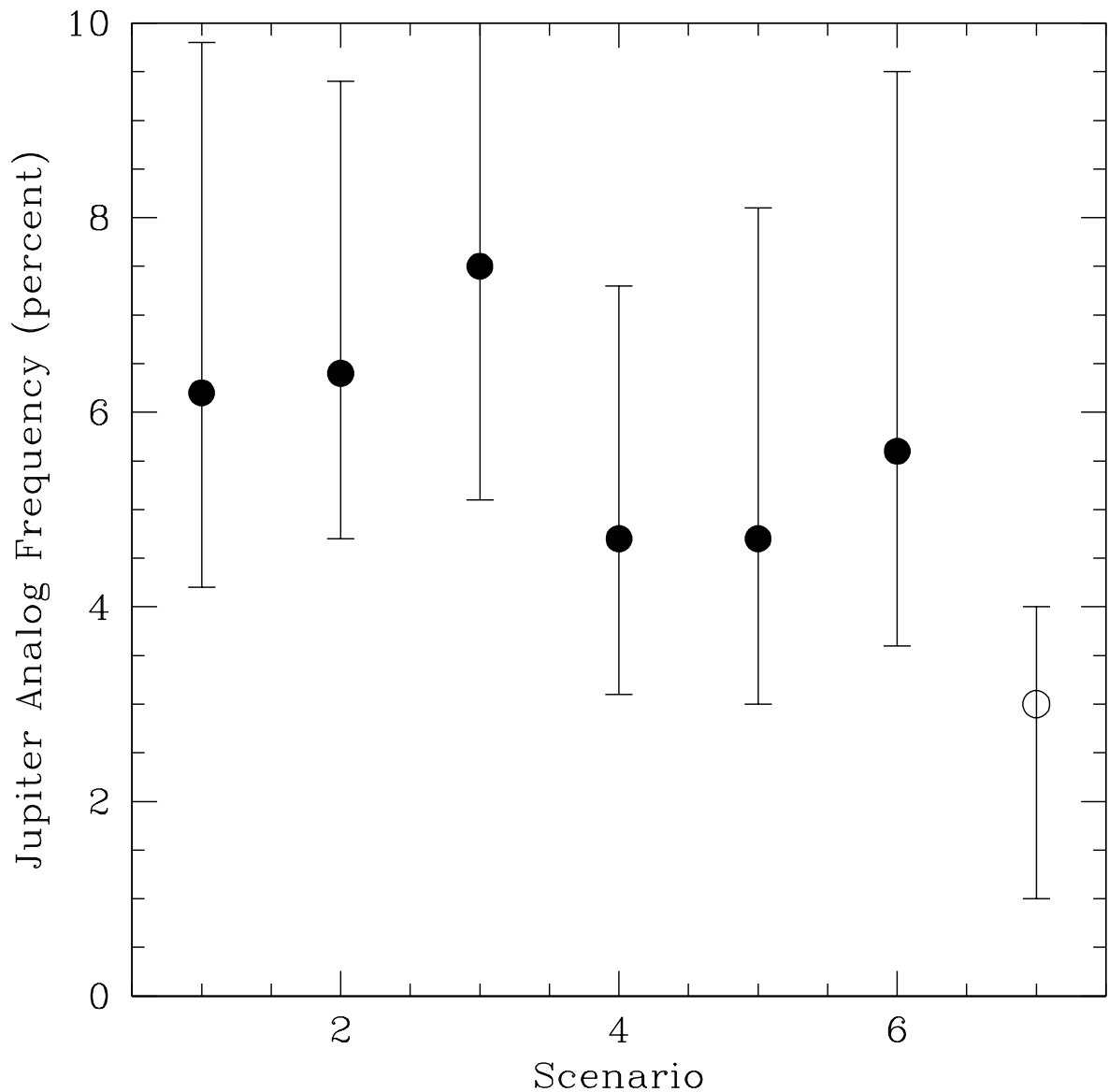}
\caption{The frequency of Jupiter analogs as computed under various 
scenarios (filled circles) as compared to the result of \citet{rowan15} 
from the Lick-Carnegie survey (open circle).  Scenarios 1, 2, and 3 use 
our definition of a Jupiter analog, while Scenarios 4, 5, and 6 use the 
R15 definition which excludes planets with $P>15$\,yr and $m>3$\,\Mjup.  
Error bars represent the 10-90\% confidence intervals.  When consistent 
definitions are used, our results are in agreement with those of R15. }
\label{scenarios}
\end{figure}

We explored this further by considering a number of different subsets of 
our results.  Figure~\ref{scenarios} shows six possibilities, all 
reported as 10-90\% confidence intervals, after R15 (as noted above).  
The six scenarios are, from left to right: (1) our adopted result using 
202 AAPS stars and our definition of Jupiter analog; (2) the same but 
using all 271 AAPS stars, including those with insufficient time 
coverage; (3) the same, but only using the 141 AAPS stars which have 
more than 40 observations; (4) using 202 AAPS stars but with the R15 
definition of Jupiter analog; (5) using all 271 AAPS stars but with the 
R15 definition of a Jupiter analog; (6) using only the 141 AAPS stars 
with $N>40$ and the R15 definition of a Jupiter analog.  As noted above, 
matching the R15 definition excludes two AAPS detections and reduces the 
derived frequency (Scenarios 4, 5, and 6).  Including all 271 AAPS 
stars, even those patently incapable of discerning these types of 
planets (too few observations, too short a baseline), we obtain the same 
underlying frequency.  Spreading the detections over more stars in the 
sample is countered by the missed-planet correction of the added stars, 
which are assumed to add no information to the detectability.  This 
leads to a larger missed planet correction as per Equation (3).  The 
effects cancel out, obviating any concern that we have somehow ``cherry 
picked'' our sample.  Similarly, by intentionally choosing the 
more-suitable stars ($N>40$, Scenarios 3 and 6), we again arrive at the 
same result but with larger uncertainties due to the smaller sample 
used.

Our AAPS sample contains 45 stars (22\%) with linear trends or 
unconstrained long-period objects imposing some curvature on the radial 
velocities.  These objects may be gas giant planets, or low-mass stars 
on orbits of thousands of years.  Direct imaging campaigns are revealing 
a population of super-Jupiter-mass objects in orbits of tens of AU 
\citep[e.g.][]{kalas08, marois08, chauvin12}, well beyond the range 
reasonably detectable by radial velocity.  The handful of such objects 
now known range in orbital separation from 9--113\,AU, and in mass from 
3--10\,\Mjup, though with large uncertainties correlated with the host 
star's age \citep{rameau13, goz14}.  From non-detections in the Gemini 
NICI planet-finding campaign, \citet{nielsen13} estimated that no more 
than 20\% of B and A stars can host planets $M>4\,\Mjup$ between 
59-460\,AU.  To date, high-contrast imaging studies favour A/B type 
stars, while radial velocity surveys traditionally prioritise Solar-type 
FGK stars.  The two techniques are looking for the same types of 
objects, but there remain these gaps in both host-star type (mass) and 
orbital separation due to the selection biases intrinsic to each 
technique.  There remains no substitute for time as we seek to elucidate 
the properties of large-separation giant planets.  Radial velocity 
surveys such as the AAPS will probe toward true Saturn analogs, and 
imaging camnpaigns will reach working angles closer to their host stars, 
toward Jupiter-like separations ($\sim$5\,AU).  Furthermore, the next 
generation of microlensing surveys \citep[e.g.][]{lee15} will make 
important contributions, being unfettered by the biases inherent to the 
former two methods.

\acknowledgements

JH is supported by USQ's Strategic Research Fund: the STARWINDS project.  
CGT is supported by Australian Research Council grants DP0774000 and 
DP130102695.  This research has made use of NASA's Astrophysics Data 
System (ADS), and the SIMBAD database, operated at CDS, Strasbourg, 
France.  This research has also made use of the Exoplanet Orbit Database 
and the Exoplanet Data Explorer at exoplanets.org \citep{wright11}.


\begin{deluxetable}{lr@{$\pm$}lr@{$\pm$}lr@{$\pm$}lr@{$\pm$}lr@{$\pm$}lr@{$\pm$}
lr@{$\pm$}ll}
\rotate
\tabletypesize{\scriptsize}
\tablecolumns{9}
\tablewidth{0pt}
\tablecaption{Jupiter Analogs from the AAPS Sample\tablenotemark{a} }
\tablehead{
\colhead{Planet} & \multicolumn{2}{c}{Period} & \multicolumn{2}{c}{$T_0$}
&
\multicolumn{2}{c}{$e$} & \multicolumn{2}{c}{$\omega$} &
\multicolumn{2}{c}{K } & \multicolumn{2}{c}{M sin $i$ } &
\multicolumn{2}{c}{$a$ } & \colhead{Reference} \\
\colhead{} & \multicolumn{2}{c}{(days)} & \multicolumn{2}{c}{(BJD-2400000)}
&
\multicolumn{2}{c}{} &
\multicolumn{2}{c}{(degrees)} & \multicolumn{2}{c}{(\ms)} &
\multicolumn{2}{c}{(\Mjup)} & \multicolumn{2}{c}{(AU)} & \colhead{}
 }
\startdata
\label{jupiters}   
HD 142 c & 6444 & 144 & 50624 & 168 & 0.32 & 0.05 & 283 & 8 & 60.7 & 3.7 
& 6.03 & 0.48 & 7.27 & 0.19 & \citet{142paper} \\
HD 70642 b & 2167 & 21 & 51853 & 177 & 0.068 & 0.039 & 295 & 29 &
27.8 & 1.1 & 1.82 & 0.11 & 3.33 & 0.05 & \citet{carter03} \\
HD 30177 b & 2514.5 & 7.7 & 51388 & 19 & 0.21 & 0.02 & 21 & 3 & 133.7 & 
1.9 & 8.4 & 0.4 & 3.57 & 0.07 & \citet{butler06} \\
HD 114613 b & 3827 & 105 & 55550.3 & (fixed) & 0.25 & 0.08 & 244 & 5 &
5.52 & 0.40 & 0.48 & 0.04 & 5.16 & 0.13 & \citet{2jupiters} \\
HD 134987 c & 5000 & 400 & 51100 & 600 & 0.12 & 0.02 & 195 & 48 &
9.3 & 0.3 & 0.82 & 0.03 & 5.8 & 0.5 & \citet{jones10} \\
HD 154857 c & 3452 & 105 & 55219 & 375 & 0.06 & 0.05 & 352 & 37 &
24.2 & 1.1 & 2.58 & 0.16 & 5.36 & 0.09 & \citet{2jupiters} \\
HD 160691 c & 4163 & 99 & 52513 & 62 & 0.029 & 0.024 & 23 & 48 &
23.2 & 0.5 & 2.00 & 0.10 & 5.3 & 0.1 & \citet{mccarthy04} \\
GJ 832 b & 3657 & 104 & 54194 & 197 & 0.08 & 0.06 & 246 & 22 & 
15.4 & 0.7 & 0.68 & 0.09 & 3.56 & 0.28 & \citet{gj832} \\
\enddata
\tablenotetext{a}{Those 202 stars with $N>30$ and $T_{obs}>$8~yr.}
\end{deluxetable}


\begin{deluxetable}{lrr}
\tabletypesize{\scriptsize}
\tablecolumns{3}
\tablewidth{0pt}
\tablecaption{Summary of Radial-Velocity Data }
\tablehead{
\colhead{Star} & \colhead{$N$} & \colhead{RMS\tablenotemark{a}}\\
\colhead{} & \colhead{} & \colhead{(\ms)}
 }
\startdata
\label{rvdata1}
GJ 832 & 109 & 3.6\tablenotemark{a} \\
GJ 729 & 30 & 20.7 \\
HD 100623 & 104 & 3.5\tablenotemark{a} \\
HD 101581 & 33 & 4.0 \\
HD 10180 & 38 & 7.2 \\
HD 101959 & 49 & 6.6 \\
HD 102117 & 63 & 4.4\tablenotemark{a} \\
HD 102365 & 187 & 2.7\tablenotemark{a} \\
HD 102438 & 59 & 4.6 \\
HD 10360 & 65 & 4.8\tablenotemark{a} \\
HD 10361 & 66 & 4.4\tablenotemark{a} \\
HD 105328 & 53 & 6.5 \\
HD 106453 & 37 & 7.6\tablenotemark{a} \\
HD 10647 & 53 & 10.6\tablenotemark{a} \\
HD 10700 & 258 & 3.5 \\
HD 107692 & 48 & 13.7 \\
HD 108147 & 58 & 14.1\tablenotemark{a} \\
HD 108309 & 69 & 4.1\tablenotemark{a} \\
HD 109200 & 39 & 4.5 \\
HD 110810 & 39 & 24.1 \\
HD 11112 & 41 & 9.3\tablenotemark{a} \\
HD 114613 & 244 & 4.0\tablenotemark{a} \\
HD 114853 & 58 & 6.9 \\
HD 115585 & 31 & 3.6 \\
HD 115617 & 247 & 2.5\tablenotemark{a} \\
HD 117105 & 32 & 6.7 \\
HD 117618 & 78 & 6.2\tablenotemark{a} \\
HD 117939 & 35 & 5.8\tablenotemark{a} \\
HD 118972 & 51 & 21.0 \\
HD 120237 & 57 & 10.9 \\
HD 122862 & 104 & 4.8 \\
HD 124584 & 44 & 5.8 \\
HD 125072 & 86 & 5.5 \\
HD 125881 & 36 & 9.3 \\
HD 128620 & 102 & 3.5\tablenotemark{a} \\
HD 128621 & 119 & 3.7\tablenotemark{a} \\
HD 128674 & 31 & 4.9 \\
HD 129060 & 45 & 36.9 \\
HD 134060 & 98 & 6.6 \\
HD 134330 & 46 & 5.7\tablenotemark{a} \\
HD 134331 & 60 & 5.5\tablenotemark{a} \\
HD 13445 & 72 & 6.1\tablenotemark{a} \\
HD 134606 & 66 & 5.4\tablenotemark{a} \\
HD 134987 & 77 & 3.4\tablenotemark{a} \\
HD 136352 & 169 & 4.7 \\
HD 140785 & 40 & 7.1 \\
HD 140901 & 117 & 11.4\tablenotemark{a} \\
HD 142 & 92 & 10.8\tablenotemark{a} \\
HD 143114 & 41 & 6.8 \\
HD 144009 & 38 & 4.4 \\
HD 144628 & 56 & 4.2 \\
HD 145417 & 34 & 10.0 \\
HD 146233 & 81 & 5.7\tablenotemark{a} \\
HD 146481 & 33 & 5.6 \\
HD 147722 & 66 & 16.9\tablenotemark{a} \\
HD 147723 & 72 & 9.6\tablenotemark{a} \\
HD 149612 & 35 & 6.5 \\
HD 150474 & 47 & 5.3 \\
HD 151337 & 47 & 6.2 \\
HD 153075 & 38 & 4.8 \\
HD 154577 & 43 & 4.2 \\
HD 154857 & 45 & 3.6\tablenotemark{a} \\
HD 155918 & 34 & 4.3 \\
HD 155974 & 50 & 9.6 \\
HD 156274B & 96 & 6.5\tablenotemark{a} \\
HD 1581 & 117 & 3.7 \\
HD 159868 & 86 & 5.8\tablenotemark{a} \\
HD 160691 & 180 & 2.4\tablenotemark{a} \\
HD 161050 & 31 & 7.6\tablenotemark{a} \\
HD 161612 & 52 & 4.5 \\
HD 163272 & 40 & 7.0 \\
HD 16417 & 121 & 3.9\tablenotemark{a} \\
HD 164427 & 44 & 6.2\tablenotemark{a} \\
HD 165269 & 33 & 13.5 \\
HD 166553 & 43 & 23.4\tablenotemark{a} \\
HD 168060 & 48 & 5.8 \\
HD 168871 & 73 & 5.0 \\
HD 17051 & 37 & 18.2\tablenotemark{a} \\
HD 172051 & 63 & 3.8 \\
HD 177565 & 106 & 4.0 \\
HD 179949 & 66 & 10.9\tablenotemark{a} \\
HD 181428 & 45 & 8.6 \\
HD 183877 & 45 & 6.0 \\
HD 187085 & 74 & 6.1\tablenotemark{a} \\
HD 189567 & 94 & 6.1 \\
HD 190248 & 235 & 4.0\tablenotemark{a} \\
HD 191408 & 185 & 3.9\tablenotemark{a} \\
HD 191849 & 42 & 8.9 \\
HD 192310 & 161 & 3.1\tablenotemark{a} \\
HD 192865 & 45 & 10.5 \\
HD 193193 & 54 & 5.6\tablenotemark{a} \\
HD 193307 & 83 & 4.5 \\
HD 194640 & 83 & 4.7 \\
HD 196050 & 56 & 7.8\tablenotemark{a} \\
HD 196068 & 35 & 6.6\tablenotemark{a} \\
HD 19632 & 30 & 24.8 \\
HD 196378 & 51 & 7.1 \\
HD 196761 & 49 & 6.5 \\
HD 196800 & 41 & 6.7 \\
HD 199190 & 55 & 4.9 \\
HD 199288 & 83 & 5.2 \\
HD 199509 & 33 & 4.7\tablenotemark{a} \\
HD 20029 & 36 & 9.8 \\
HD 20201 & 35 & 9.3\tablenotemark{a} \\
HD 202560 & 47 & 5.0 \\
HD 202628 & 30 & 10.9 \\
HD 2039 & 46 & 14.0\tablenotemark{a} \\
HD 204287 & 48 & 5.2\tablenotemark{a} \\
HD 204385 & 43 & 6.6 \\
HD 205390 & 36 & 9.6 \\
HD 206395 & 45 & 18.5 \\
HD 207129 & 124 & 5.2 \\
HD 20766 & 57 & 6.7\tablenotemark{a} \\
HD 207700 & 36 & 5.0\tablenotemark{a} \\
HD 20782 & 57 & 6.1\tablenotemark{a} \\
HD 20794 & 145 & 3.6 \\
HD 20807 & 99 & 5.0 \\
HD 208487 & 49 & 8.2\tablenotemark{a} \\
HD 208998 & 36 & 8.3\tablenotemark{a} \\
HD 209268 & 30 & 5.5 \\
HD 209653 & 42 & 5.3 \\
HD 210918 & 72 & 6.1 \\
HD 211317 & 44 & 5.3 \\
HD 211998 & 47 & 7.4 \\
HD 212168 & 51 & 5.6 \\
HD 212330 & 33 & 3.8\tablenotemark{a} \\
HD 212708 & 37 & 4.5\tablenotemark{a} \\
HD 213240 & 37 & 5.5\tablenotemark{a} \\
HD 214759 & 32 & 7.1 \\
HD 214953 & 84 & 4.9\tablenotemark{a} \\
HD 2151 & 172 & 4.7\tablenotemark{a} \\
HD 216435 & 78 & 7.1\tablenotemark{a} \\
HD 216437 & 56 & 4.7\tablenotemark{a} \\
HD 217958 & 37 & 8.2 \\
HD 217987 & 42 & 6.0 \\
HD 219077 & 72 & 4.9\tablenotemark{a} \\
HD 22104 & 43 & 10.9 \\
HD 221420 & 86 & 3.9\tablenotemark{a} \\
HD 222237 & 34 & 5.6\tablenotemark{a} \\
HD 222335 & 30 & 4.5 \\
HD 222480 & 33 & 7.4\tablenotemark{a} \\
HD 222668 & 36 & 5.8 \\
HD 223171 & 63 & 5.8\tablenotemark{a} \\
HD 225213 & 35 & 5.4 \\
HD 23079 & 40 & 6.6\tablenotemark{a} \\
HD 23127 & 44 & 11.6\tablenotemark{a} \\
HD 23249 & 93 & 3.2 \\
HD 26965 & 111 & 4.5\tablenotemark{a} \\
HD 27274 & 30 & 6.9 \\
HD 27442 & 103 & 7.3\tablenotemark{a} \\
HD 28255A & 68 & 7.4\tablenotemark{a} \\
HD 28255B & 44 & 15.2\tablenotemark{a} \\
HD 30177 & 41 & 9.1\tablenotemark{a} \\
HD 30295 & 33 & 9.2 \\
HD 30876 & 32 & 7.4\tablenotemark{a} \\
HD 31527 & 32 & 6.4 \\
HD 31827 & 31 & 8.8 \\
HD 36108 & 39 & 4.2 \\
HD 3823 & 81 & 5.6 \\
HD 38283 & 66 & 4.6\tablenotemark{a} \\
HD 38382 & 47 & 5.9 \\
HD 38973 & 48 & 5.2 \\
HD 39091 & 75 & 5.4\tablenotemark{a} \\
HD 4308 & 115 & 4.3\tablenotemark{a} \\
HD 43834 & 138 & 5.5\tablenotemark{a} \\
HD 44120 & 42 & 3.8 \\
HD 44447 & 38 & 5.7\tablenotemark{a} \\
HD 44594 & 45 & 5.8\tablenotemark{a} \\
HD 45289 & 36 & 7.4\tablenotemark{a} \\
HD 45701 & 35 & 5.5\tablenotemark{a} \\
HD 53705 & 138 & 4.4 \\
HD 53706 & 48 & 3.5 \\
HD 55693 & 41 & 6.5 \\
HD 55720 & 30 & 3.9 \\
HD 59468 & 47 & 4.9 \\
HD 65907A & 75 & 6.1 \\
HD 67199 & 49 & 6.6\tablenotemark{a} \\
HD 67556 & 30 & 14.3 \\
HD 69655 & 30 & 5.6 \\
HD 70642 & 49 & 4.6\tablenotemark{a} \\
HD 70889 & 40 & 16.1 \\
HD 72673 & 63 & 3.0 \\
HD 72769 & 31 & 3.6\tablenotemark{a} \\
HD 73121 & 44 & 5.8 \\
HD 73524 & 85 & 5.4 \\
HD 74868 & 60 & 7.6 \\
HD 75289 & 49 & 6.4\tablenotemark{a} \\
HD 7570 & 57 & 6.2 \\
HD 76700 & 43 & 6.4\tablenotemark{a} \\
HD 78429 & 38 & 8.6 \\
HD 80913 & 35 & 11.3\tablenotemark{a} \\
HD 83529A & 32 & 4.7\tablenotemark{a} \\
HD 84117 & 145 & 5.6 \\
HD 85512 & 31 & 5.0 \\
HD 85683 & 30 & 8.8 \\
HD 86819 & 36 & 8.7\tablenotemark{a} \\
HD 88742 & 36 & 10.1\tablenotemark{a} \\
HD 9280 & 33 & 10.3\tablenotemark{a} \\
HD 92987 & 52 & 5.3\tablenotemark{a} \\
HD 93385 & 46 & 6.9\tablenotemark{a} \\
HD 96423 & 42 & 5.3 \\
\enddata
\tablenotetext{a}{Velocity scatter after removal of known planets and trends.}
\end{deluxetable}

\begin{deluxetable}{lr@{$\pm$}lr@{$\pm$}lr@{$\pm$}lr@{$\pm$}lr@{$\pm$}lr@{$\pm$}l}
\tabletypesize{\scriptsize}
\tablecolumns{7}
\tablewidth{0pt}
\tablecaption{Summary of Detection Limits: Mean velocity amplitude $K$ 
detectable for orbital periods 1000-6000 days }
\tablehead{
\colhead{Star} & \multicolumn{10}{c}{Recovery Rate (percent)} \\
\colhead{} & \multicolumn{2}{c}{99} & \multicolumn{2}{c}{90} & \multicolumn{2}{c}{70} & \multicolumn{2}{c}{50} & 
\multicolumn{2}{c}{30} & \multicolumn{2}{c}{10}
 }
\startdata
\label{meankays}
GJ 832    & 3.5 & 1.3 & 2.9 & 1.2 & 2.1 & 0.9 & 1.5 & 0.2 & 1.4 & 0.1 & 1.2 & 0.1 \\
GJ 729    & 80.6 & 43.6 & 79.0 & 39.9 & 73.5 & 29.9 & 61.6 & 18.9 & 53.1 & 14.2 & 45.1 & 12.5 \\
HD 100623 & 4.5 & 0.9 & 4.2 & 0.8 & 3.5 & 0.7 & 3.1 & 0.7 & 2.7 & 0.8 & 2.5 & 0.8 \\
HD 101581 & 15.6 & 6.7 & 13.4 & 2.5 & 11.1 & 1.3 & 9.5 & 1.5 & 8.3 & 1.8 & 7.0 & 1.9 \\
HD 10180  & 18.7 & 4.1 & 17.3 & 2.2 & 15.2 & 1.2 & 13.6 & 0.8 & 12.1 & 1.0 & 10.7 & 1.1 \\
HD 101959 & 12.6 & 1.8 & 11.9 & 1.0 & 10.8 & 0.6 & 10.1 & 0.5 & 9.1 & 0.8 & 8.2 & 1.2 \\
HD 102117 & 6.0 & 2.4 & 5.4 & 0.4 & 5.0 & 0.3 & 4.5 & 0.4 & 3.8 & 0.4 & 3.3 & 0.4 \\
HD 102365 & 2.2 & 0.3 & 2.0 & 0.2 & 1.7 & 0.2 & 1.5 & 0.2 & 1.2 & 0.2 & 1.1 & 0.1 \\
HD 102438 & 5.7 & 0.4 & 5.4 & 0.3 & 4.7 & 0.4 & 4.3 & 0.4 & 3.8 & 0.4 & 3.4 & 0.5 \\
HD 10360  & 10.0 & 1.3 & 9.0 & 0.9 & 7.6 & 0.7 & 6.2 & 1.4 & 4.3 & 2.2 & 3.3 & 1.8 \\
HD 10361  & 7.5 & 0.6 & 6.9 & 0.6 & 5.5 & 0.5 & 4.8 & 0.6 & 4.2 & 0.7 & 3.7 & 0.7 \\
HD 105328 & 10.2 & 0.9 & 9.7 & 0.7 &  8.1 & 0.7 &  6.6 & 1.3 & 5.4 & 1.7 & 4.7 & 1.8 \\
HD 106453 & 19.6 & 4.4 & 18.1 & 3.1 & 15.3 & 1.9 & 13.2 & 1.7 & 13.5 & 7.3 & 10.7 & 4.2 \\
HD 10647  & 22.9 & 5.6 & 18.9 & 2.2 & 15.3 & 1.4 & 13.4 & 1.4 & 11.3 & 1.6 &  9.0 & 2.0 \\
HD 10700  & 2.9 & 0.8 & 2.7 & 0.8 & 2.0 & 0.6 & 1.5 & 0.4 & 1.3 & 0.3 & 1.1 & 0.1 \\
HD 107692 & 26.3 & 2.6 & 24.4 & 2.2 & 19.5 & 1.8 & 16.8 & 1.5 & 14.8 & 1.4 & 13.2 & 1.3 \\
HD 108309 & 6.7 & 0.7 & 6.2 & 0.6 & 5.1 & 0.4 & 4.4 & 0.6 & 3.6 & 0.8 & 3.1 & 0.7 \\
HD 109200 & 10.7 & 1.1 & 10.1 & 0.6 & 9.2 & 0.5 & 8.7 & 0.4 & 7.8 & 0.6 &  6.9 & 0.6 \\
HD 110810 & 72.7 & 23.3 & 68.2 & 18.5 & 64.1 & 18.4 & 51.8 & 12.2 & 43.3 & 7.0 & 36.9 & 4.0 \\
HD 11112  & 25.8 & 4.6 & 23.7 & 2.4 & 19.9 & 1.6 & 16.2 & 1.9 & 13.7 & 2.1 & 11.7 & 2.4 \\
HD 114613 & 3.6 & 0.4 & 3.4 & 0.3 & 2.8 & 0.4 & 2.4 & 0.3 & 2.0 & 0.5 & 1.9 & 0.5 \\
HD 114853 & 10.7 & 1.1 &  9.7 & 1.9 & 7.8 & 1.0 & 6.3 & 1.3 &  5.1 & 1.5 & 4.4 & 1.5 \\
HD 115585 & 15.3 & 6.2 & 14.2 & 4.6 & 13.0 & 2.6 & 11.1 & 2.3 & 9.8 & 2.4 & 8.5 & 2.3 \\
HD 115617 & 2.6 & 0.5 & 2.5 & 0.5 & 2.2 & 0.4 & 2.1 & 0.4 & 1.9 & 0.3 & 1.8 & 0.3 \\
HD 117105 & 27.6 & 15.0 & 25.7 & 9.9 & 25.5 & 7.7 & 19.8 & 4.0 & 17.0 & 4.1 & 14.3 & 4.1 \\
HD 117618 & 8.5 & 1.4 & 7.9 & 1.0 &  6.5 & 0.7 &  5.5 & 0.8 & 4.6 & 0.7 & 4.0 & 0.5 \\
HD 117939 & 19.1 & 3.8 & 17.8 & 2.7 & 15.6 & 2.1 & 13.2 & 2.4 & 13.0 & 1.9 & 11.4 & 1.7 \\
HD 118972 & 46.0 & 10.0 & 41.2 & 6.4 & 32.7 & 4.0 & 29.5 & 3.4 & 25.6 & 4.0 & 20.6 & 3.7 \\
HD 120237 & 18.4 & 1.7 & 17.6 & 1.4 & 15.5 & 1.0 & 13.5 & 1.7 & 11.7 & 2.1 & 10.5 & 2.1 \\
HD 122862 & 6.0 & 2.3 & 5.3 & 0.7 & 4.2 & 0.5 & 3.4 & 0.4 & 2.7 & 0.4 & 2.3 & 0.5 \\
HD 124584 & 12.6 & 1.9 & 11.6 & 1.0 & 10.4 & 0.7 & 9.5 & 0.5 & 8.5 & 0.6 & 7.4 & 0.6 \\
HD 125072 & 6.9 & 3.6 & 6.3 & 3.2 &  4.9 & 2.8 & 4.7 & 2.5 & 3.4 & 2.0 &  2.9 & 1.7 \\
HD 125881 & 32.2 & 6.7 & 29.0 & 3.7 &  24.8 & 2.7 & 21.3 & 2.9 & 17.2 & 2.9 & 14.1 & 2.3 \\
HD 128620 &  8.4 & 6.2 & 5.7 & 2.6 & 3.8 & 2.3 & 2.8 & 1.5 & 1.7 & 1.0 & 1.4 & 0.7 \\
HD 128621 & 7.2 & 3.0 & 6.7 & 2.6 &  6.1 & 2.4 & 5.3 & 1.7 & 4.7 & 1.6 & 3.0 & 0.7 \\
HD 128674 & 19.2 & 9.3 & 17.5 & 6.8 & 16.1 & 4.6 & 12.6 & 2.2 & 10.9 & 2.3 &  9.3 & 2.1 \\
HD 129060 & 87.4 & 7.3 & 81.1 & 6.9 & 70.7 & 6.8 & 63.4 & 7.5 & 54.5 & 9.0 & 46.8 & 9.2 \\
HD 134060 & 8.2 & 3.1 & 7.1 & 0.8 & 5.8 & 0.6 & 4.6 & 1.0 & 3.2 & 0.9 & 2.3 & 0.6 \\
HD 134330 & 12.5 & 1.8 & 11.4 & 0.8 & 9.9 & 0.7 & 8.7 & 1.0 & 7.0 & 1.6 & 5.9 & 1.6 \\
HD 134331 & 8.6 & 0.5 &  8.3 & 0.5 & 7.5 & 0.4 & 6.8 & 0.5 & 5.9 & 0.6 & 5.1 & 0.7 \\
HD 13445  &  7.2 & 1.4 &  6.5 & 0.9 & 5.4 & 0.7 & 4.5 & 1.0 & 3.6 & 1.1 &  2.8 & 0.9 \\
HD 134606 & 8.2 & 1.9 & 7.4 & 1.0 & 5.7 & 1.1 & 4.9 & 1.7 & 3.3 & 1.5 & 2.8 & 1.5 \\
HD 134987 &  5.0 & 1.6 & 4.3 & 0.5 &  3.7 & 0.3 &  3.3 & 0.2 & 2.8 & 0.4 & 2.4 & 0.5 \\
HD 136352 & 6.7 & 1.6 & 6.0 & 0.9 & 5.0 & 0.6 & 4.4 & 0.5 & 3.6 & 0.5 & 3.1 & 0.5 \\
HD 140785 & 14.1 & 2.8 & 13.1 & 2.0 & 11.5 & 1.3 & 10.3 & 0.9 & 9.0 & 0.7 &  7.9 & 0.7 \\
HD 140901 & 20.2 & 4.2 &  17.0 & 3.0 & 11.1 & 2.8 & 8.7 & 3.4 & 4.4 & 2.6 & 3.5 & 2.2 \\
HD 142    & 14.1 & 2.5  & 12.5 & 1.2 & 10.7 & 0.9 & 9.5 & 0.8 & 8.2 & 0.9 & 7.4 & 1.1 \\
HD 143114 & 16.6 & 3.2  & 15.4 & 1.8 &  13.6 & 1.1 & 12.1 & 1.3 & 11.0 & 1.4 &  10.0 & 1.5  \\
HD 144009 & 12.3 & 2.4 & 11.2 & 1.5 & 9.7 & 1.0 & 8.8 & 1.2 & 7.6 & 1.7  &  6.6 & 2.0 \\
HD 144628 & 6.9 & 0.8 & 6.6 & 0.7 & 5.7 & 0.6 & 4.9 & 0.8 & 4.1 & 1.0 & 3.6 & 1.0 \\
HD 145417 & 36.0 & 11.4 & 31.3 & 8.0 & 24.9 & 3.5 & 21.7 & 2.6 & 19.3 & 2.7 &  17.1 & 2.9 \\
HD 146233 & 8.7 & 6.1 &  8.0 & 5.6 &  5.9 & 4.0 & 4.2 & 2.6 & 3.2 & 1.8 & 2.6 & 1.3 \\
HD 146481 &  17.3 & 3.7 & 15.9 & 2.6 & 13.7 & 1.5 & 12.2 & 1.3 & 10.7 & 1.9 & 9.6 & 1.9  \\
HD 147722 & 23.1 & 3.0 & 21.6 & 2.5 & 18.8 & 1.3 & 18.4 & 3.2 & 15.0 & 1.9 & 12.9 & 2.2 \\
HD 147723 & 11.5 & 1.1 & 11.0 & 1.0  & 9.9 & 0.4 & 10.0 & 1.7 & 8.3 & 0.5 & 7.4 & 0.7 \\
HD 149612 & 15.8 & 2.6 & 14.2 & 2.0 & 12.0 & 1.4 & 10.5 & 1.1 & 9.4 & 1.3 & 8.1 & 1.2 \\
HD 150474 & 11.5 & 1.0 & 10.9 & 0.9 & 9.7 & 0.6 & 8.5 & 0.8 & 6.7 & 1.3 & 5.7 & 1.5 \\
HD 151337 & 11.4 & 1.3 & 10.7 & 0.9 & 9.7 & 0.4 & 8.9 & 0.6 & 8.1 & 0.7 &  7.1 & 0.6 \\
HD 153075 & 13.7 & 2.1 & 12.6 & 1.6 & 10.8 & 1.0 & 9.9 & 0.8 & 9.1 & 0.9 & 8.2 & 1.0 \\
HD 154577 & 8.4 & 0.5 & 8.1 & 0.5 & 7.3 & 0.3 & 6.8 & 0.4 &  6.3 & 0.6 & 5.9 & 0.7 \\
HD 154857 & 8.6 & 1.1 &  7.8 & 0.9 & 6.6 & 0.6 & 5.9 & 0.4 & 5.5 & 0.4 & 4.9 & 0.5 \\
HD 155918 & 15.9 & 7.4 & 13.7 & 3.7 & 11.9 & 2.2 &  9.8 & 1.8 & 9.3 & 1.3 & 8.1 & 1.4 \\
HD 155974 & 17.5 & 2.0 & 16.1 & 1.2 & 14.0 & 0.8 & 12.9 & 0.7 & 11.7 & 0.8 & 10.3 & 1.0 \\
HD 156274B & 9.2 & 4.5 & 7.4 & 2.6 & 5.5 & 1.4 & 4.1 & 1.4 & 2.3 & 1.4 & 1.8 & 0.9  \\
HD 1581   & 4.1 & 0.9 & 3.7 & 0.7 &  2.9 & 0.4 & 2.5 & 0.3 & 2.1 & 0.3  & 1.9 & 0.2 \\
HD 159868 &  7.7 & 0.8 &  7.4 & 0.8 & 6.9 & 0.7  &  6.2 & 0.9 & 5.2 & 1.3 &  4.5 & 1.4 \\
HD 160691 & 2.0 & 0.2 & 1.8 & 0.2 & 1.6 & 0.1 & 1.4 & 0.1 &  1.3 & 0.1 & 1.2 & 0.1 \\
HD 161050 & 45.0 & 22.6 & 44.5 & 17.2 & 36.0 & 7.3  & 29.0 & 5.7  & 24.2 & 6.6 &  19.0 & 7.7 \\
HD 161612 & 7.4 & 3.1 & 6.6 & 0.4 & 5.8 & 0.4 & 5.2 & 0.3 & 4.6 & 0.4 & 4.1 & 0.6 \\
HD 163272 & 19.0 & 2.6 & 17.9 & 2.2 & 15.1 & 1.0 & 13.4 & 1.0 & 12.0 & 1.2 & 10.3 & 1.5 \\
HD 16417  & 5.7 & 2.4 & 4.5 & 0.6 & 3.7 & 0.4 & 3.3 & 0.6 & 2.6 & 0.8 & 2.0 & 0.6 \\
HD 164427 & 15.9 & 7.2  & 12.8 & 1.9 & 10.6 & 1.1 &  9.3 & 1.0 & 8.1 & 1.3 &  6.8 & 1.2 \\
HD 165269 & 48.3 & 23.7 & 46.4 & 18.8 &  35.6 & 8.1 & 30.0 & 6.3 & 22.8 & 7.8 & 17.6 & 7.1 \\
HD 166553 & 54.3 & 10.8 & 52.0 & 9.9 &  46.6 & 5.1 & 41.6 & 3.3 & 37.8 & 3.5 & 32.7 & 4.1 \\
HD 168060 & 10.4 & 2.6 & 9.4 & 1.4 & 8.0 & 1.0 & 7.0 & 0.7 &  6.2 & 0.4 & 5.4 & 0.4 \\
HD 168871 & 7.5 & 1.4 &  6.8 & 1.0 & 5.4 & 0.7 & 4.6 & 0.7 &  4.0 & 0.9 &  3.5 & 0.8 \\
HD 17051  &  55.3 & 14.5 &  51.0 & 10.3 & 42.3 & 4.4 &  36.2 & 3.1 & 31.6 & 3.3 & 26.8 & 3.2 \\
HD 172051 & 10.6 & 10.2 & 7.8 & 2.7 &  5.9 & 0.6 &  5.2 & 0.4 & 4.2 & 1.0 & 3.0 & 0.9 \\
HD 177565 & 6.8 & 5.9 & 6.0 & 4.3 &  4.7 & 3.0 & 3.9 & 2.2 &  3.4 & 1.8 & 2.6 & 1.3 \\
HD 179949 & 19.1 & 6.0 & 17.6 & 5.2 & 14.2 & 3.4  & 12.6 & 5.6 & 8.8 & 4.1  &  6.4 & 2.9 \\
HD 181428 & 16.5 & 4.2 & 14.6 & 1.2 & 13.0 & 0.6 & 11.7 & 0.7 & 10.4 & 0.9 & 9.1 & 1.1  \\
HD 183877 & 11.8 & 1.1 & 11.1 & 0.7 & 9.7 & 0.6 &  8.9 & 0.6 &  8.0 & 1.4 & 7.2 & 1.8 \\
HD 187085 & 7.4 & 0.6  & 7.1 & 0.4  & 6.6 & 0.3 & 6.2 & 0.3 & 5.6 & 0.4 & 5.0 & 0.5 \\
HD 189567 & 7.5 & 1.0  & 6.6 & 0.6  &  5.5 & 0.8 & 4.2 & 1.2  & 3.0 & 1.3  & 2.2 & 0.9 \\
HD 190248 & 4.8 & 0.9 &  4.5 & 0.8  & 4.0 & 0.7  & 3.5 & 0.3  &  3.1 & 0.3 & 2.8 & 0.4 \\
HD 191408 & 4.9 & 0.9 & 4.2 & 0.4 & 3.5 & 0.5 & 3.0 & 0.4 & 2.4 & 0.4 & 2.0 & 0.4 \\
HD 191849 & 16.5 & 5.9 &  14.7 & 2.1 &  12.6 & 1.1 & 11.3 & 1.1 & 9.9 & 1.2 & 8.8 & 1.5 \\
HD 192310 & 2.7 & 0.4 & 2.5 & 0.3 & 2.1 & 0.2 & 1.9 & 0.2 &  1.8 & 0.2  & 1.6 & 0.2 \\
HD 192865 & 23.3 & 2.7 & 21.4 & 2.1 &  18.0 & 2.1 & 15.4 & 1.4 & 13.4 & 1.4 & 11.4 & 1.6 \\
HD 193193 &  9.8 & 0.9 &  9.2 & 0.6 & 8.0 & 0.6 & 7.5 & 0.6 &  6.8 & 0.7 & 6.1 & 0.6 \\
HD 193307 & 7.4 & 1.8 & 6.5 & 1.1 & 5.6 & 0.6 &  4.8 & 0.6 & 3.9 & 0.6 & 3.1 & 0.4 \\
HD 194640 & 7.7 & 4.6 &  6.3 & 0.9 & 5.4 & 0.3 & 4.8 & 0.4 & 4.1 & 0.6 &  3.3 & 0.6 \\
HD 196050 & 12.5 & 2.1 & 11.9 & 1.8 & 10.6 & 1.1 & 9.7 & 0.9 & 8.8 & 0.7 &  7.8 & 0.7 \\
HD 196068 & 21.2 & 2.7  & 20.1 & 2.0 &  18.0 & 1.2 & 15.1 & 2.6 & 14.5 & 1.3 & 12.6 & 1.7 \\
HD 19632  & 82.4 & 43.0 & 86.8 & 44.2 & 96.6 & 37.3 &  79.1 & 29.7 & 61.6 & 25.4 & 44.6 & 22.7 \\
HD 196378 & 14.6 & 1.8 & 13.6 & 1.0 & 11.8 & 0.8 & 10.6 & 0.7 &  9.7 & 0.8 &  8.4 & 1.0 \\
HD 196761 & 13.3 & 3.4 & 11.2 & 1.7  & 8.5 & 1.1 & 7.2 & 1.1 &  5.7 & 0.8 & 4.3 & 1.1 \\
HD 196800 & 17.3 & 2.4 &  16.0 & 1.2 & 14.2 & 0.8 & 12.9 & 0.9 & 11.1 & 1.4 & 9.2 & 1.8 \\
HD 199190 &  8.2 & 0.7 &  7.8 & 0.5 & 6.8 & 0.4 &  6.2 & 0.4 & 5.6 & 0.3 & 5.0 & 0.4 \\
HD 199288 & 8.2 & 0.9 & 7.4 & 0.8 &  5.8 & 0.7 & 5.1 & 0.5 & 4.4 & 0.4 & 3.6 & 0.4 \\
HD 199509 & 19.9 & 5.9 & 18.5 & 5.0 & 15.4 & 2.6 & 12.7 & 1.6 & 10.9 & 1.3 &  9.6 & 1.5 \\
HD 20029  & 42.9 & 18.4 & 35.6 & 10.5 &  27.0 & 3.0 &  23.0 & 2.0 &  20.3 & 2.2 & 17.5 & 2.8 \\
HD 20201  & 29.2 & 7.0 & 25.8 & 4.2 &  21.6 & 2.3 & 17.7 & 1.9 & 16.2 & 1.6  & 13.8 & 1.7 \\
HD 202560 & 10.9 & 0.9 & 10.3 & 0.7 & 9.2 & 0.6 &  7.9 & 1.1 & 6.2 & 1.3 & 5.0 & 1.4  \\
HD 202628 & 51.9 & 30.2 & 52.9 & 30.3 &  48.6 & 18.8 & 38.0 & 9.9 & 31.4 & 8.9  & 25.0 & 8.4 \\
HD 2039   & 34.6 & 13.7 & 26.7 & 4.8 & 20.4 & 3.2 & 17.4 & 3.4 &  15.2 & 3.4 & 13.2 & 3.5 \\
HD 204287 & 11.1 & 1.2  & 10.3 & 0.7 & 9.1 & 0.6 & 7.8 & 1.0 & 5.9 & 1.4 & 4.5 & 1.0 \\
HD 204385 & 14.8 & 1.3  & 13.8 & 1.1 & 12.0 & 1.1 & 10.7 & 1.3 & 9.5 & 2.3 & 8.4 & 2.5 \\
HD 205390 & 34.4 & 8.0  & 30.2 & 4.2  & 25.3 & 2.6 &  22.0 & 2.8 & 17.7 & 3.9  & 13.9 & 3.8 \\
HD 206395 & 39.8 & 9.8 & 34.9 & 4.5 & 30.5 & 2.3 & 27.5 & 2.2 & 24.5 & 2.1 & 21.5 & 2.3 \\
HD 207129 &  7.8 & 1.6 & 6.8 & 0.9 & 5.4 & 0.9 & 4.2 & 1.1  & 2.8 & 1.0 &  1.7 & 0.3 \\
HD 20766  & 10.9 & 0.7 & 10.5 & 0.6 & 9.5 & 0.6 &  8.7 & 0.6 &  7.9 & 0.7 & 7.1 & 0.5 \\
HD 207700 & 15.6 & 2.1 & 14.4 & 1.5 & 12.8 & 0.8 & 11.6 & 0.6 & 10.8 & 0.7  & 9.8 & 1.0 \\
HD 20782  & 10.5 & 4.0 & 9.4 & 1.2 & 8.4 & 0.6 &  7.5 & 0.6 & 6.4 & 0.9 &  5.5 & 0.9 \\
HD 20794  & 3.2 & 0.8  & 2.9 & 0.4 & 2.6 & 0.3 & 2.3 & 0.3 & 2.1 & 0.1  & 1.9 & 0.1 \\
HD 20807  & 8.7 & 7.4  & 6.3 & 3.4 &  4.4 & 2.0 & 3.1 & 1.8 &  2.5 & 1.3 & 2.0 & 0.9 \\
HD 208487 &  17.0 & 4.1 & 15.2 & 2.9 & 12.6 & 1.2 & 11.3 & 1.1 & 10.0 & 1.4 & 8.6 & 1.8 \\
HD 208998 & 20.8 & 3.9 & 19.5 & 2.4 & 18.3 & 1.6 & 16.0 & 1.4 & 14.8 & 2.0 &  12.4 & 2.4 \\
HD 209268 & 22.3 & 13.9  & 26.8 & 19.2  & 30.9 & 11.6 & 24.4 & 3.8 & 19.5 & 3.5 &  16.7 & 3.3 \\
HD 209653 & 13.7 & 1.5 & 13.0 & 1.2 & 11.6 & 0.9 & 10.6 & 0.7  & 9.7 & 0.7 &  8.5 & 0.8 \\
HD 210918 & 8.3 & 2.8 & 7.0 & 1.6 & 5.3 & 1.1 & 3.7 & 1.6 & 1.9 & 1.2 & 1.4 & 0.6  \\
HD 211317 & 13.3 & 4.1 &  11.4 & 1.2 &  9.7 & 0.8  & 8.6 & 1.1 & 7.2 & 1.6  & 5.7 & 1.5 \\
HD 211998 & 21.6 & 10.4 & 18.0 & 5.9 & 12.8 & 2.1 & 10.9 & 1.5 & 8.4 & 1.6 & 5.7 & 1.8 \\
HD 212168 & 10.9 & 1.0 & 10.2 & 0.8 & 8.4 & 0.8 & 7.1 & 0.8 & 6.1 & 0.8 &  5.2 & 0.8 \\
HD 212330 & 14.9 & 4.2 & 13.9 & 2.7 & 12.4 & 1.7 & 10.9 & 1.1 & 9.7 & 1.1 &  8.7 & 1.4 \\
HD 212708 & 15.8 & 8.3 & 11.9 & 2.5 & 9.8 & 1.1 & 8.9 & 1.7 & 8.2 & 3.7 & 5.5 & 2.5 \\
HD 213240 & 17.8 & 5.5 & 16.4 & 5.2 & 13.3 & 1.5 & 11.6 & 1.4 & 10.2 & 1.6 &  8.7 & 2.0 \\
HD 214759 & 21.8 & 8.8 & 21.4 & 6.4 & 20.0 & 3.5 & 16.2 & 2.2 & 13.4 & 2.9 &  11.3 & 3.2 \\
HD 214953 & 9.9 & 6.1  & 8.1 & 4.0  & 5.7 & 1.6 & 4.7 & 1.0 & 3.8 & 0.9 &  2.7 & 0.6 \\
HD 2151   &  7.5 & 1.7 &  7.1 & 1.5 & 6.3 & 1.3 & 5.2 & 1.0 & 4.7 & 2.0 & 7.3 & 3.1 \\
HD 216435 & 10.9 & 3.8 &  9.2 & 1.7 & 7.5 & 1.4 &  6.3 & 1.0 &  5.4 & 0.8 & 4.6 & 0.7 \\
HD 216437 & 8.2 & 1.2 &  7.6 & 0.8 & 6.5 & 0.4 &  5.9 & 0.5 &  5.4 & 0.6 &  4.9 & 0.6 \\
HD 217958 & 26.5 & 7.1  & 22.9 & 3.5 & 20.0 & 1.6 & 17.6 & 1.8 & 14.2 & 2.9 & 11.5 & 2.9 \\
HD 219077 &  8.8 & 3.7 & 7.5 & 1.8 & 6.1 & 0.7 &  5.1 & 0.6 & 4.1 & 0.6 &  3.2 & 0.5  \\
HD 22104  & 21.0 & 3.8 & 18.7 & 2.9 & 15.8 & 1.7 & 14.2 & 1.5 & 13.1 & 1.7 & 11.5 & 1.8 \\
HD 221420 & 5.4 & 0.5 &  4.9 & 0.4 & 4.3 & 0.4 &  3.8 & 0.4 & 3.0 & 0.7 &  2.4 & 0.7 \\
HD 222237 & 18.4 & 6.2 & 16.6 & 3.3 & 14.4 & 1.5 &  11.9 & 1.6 &  11.3 & 1.2 &  9.9 & 1.6 \\
HD 222335 &  24.4 & 15.1 & 26.5 & 15.9 & 23.3 & 6.8 &  18.4 & 2.3 &  16.0 & 2.2 & 13.6 & 2.3 \\
HD 222480 & 29.2 & 7.7 & 26.8 & 7.7 & 21.8 & 4.6 & 17.7 & 4.5 & 16.6 & 3.5 & 14.1 & 3.5 \\
HD 222668 & 21.9 & 10.5 & 17.6 & 5.2 &  14.0 & 2.3 &  11.8 & 1.7 &  9.8 & 1.4  &  8.1 & 0.9 \\
HD 223171 & 10.7 & 3.4 & 9.1 & 0.9 & 7.5 & 0.9 & 6.9 & 1.1 & 5.3 & 1.2 &  4.5 & 1.1 \\
HD 225213 & 25.1 & 20.0  & 23.0 & 11.7 & 16.7 & 4.1  & 15.5 & 2.8 & 13.4 & 2.4 & 11.4 & 2.2 \\
HD 23079  & 15.1 & 2.6 & 13.6 & 1.6 & 11.6 & 1.0 & 10.1 & 0.9 & 9.0 & 1.0 & 7.9 & 0.9 \\
HD 23127  & 20.4 & 2.6 & 19.6 & 6.5 &  15.8 & 2.0 & 13.8 & 2.3 & 12.1 & 2.7 & 10.5 & 2.8 \\
HD 23249  & 9.3 & 8.9 & 7.6 & 5.8 & 4.4 & 0.6 & 3.9 & 0.3 & 3.4 & 0.3 & 2.9 & 0.3 \\
HD 26965  & 8.8 & 5.1  & 7.2 & 1.7 & 6.2 & 1.2 &  5.6 & 0.9 &  5.1 & 0.7  & 4.6 & 0.7 \\
HD 27274  & 36.4 & 18.5 &  32.1 & 14.0 & 29.2 & 10.1 & 23.0 & 6.2 &  19.3 & 4.9 & 15.6 & 3.7 \\
HD 27442  & 9.1 & 1.1  & 8.5 & 1.0 &  6.9 & 0.9 &  5.2 & 1.2 & 4.1 & 1.3 & 3.5 & 1.4  \\
HD 28255A & 14.4 & 4.5 & 12.5 & 1.6 & 10.2 & 0.8 & 9.2 & 0.8 & 7.8 & 1.0 & 6.5 & 0.9 \\
HD 28255B & 32.9 & 2.8 & 30.8 & 1.4 & 27.8 & 1.1 & 25.1 & 1.7 & 22.3 & 3.2 & 20.0 & 3.3 \\
HD 31077  & 21.4 & 4.3  & 19.3 & 3.2 & 16.2 & 1.4 & 14.0 & 0.7 &  12.7 & 1.1 &  10.9 & 1.4 \\
HD 30295  &  31.2 & 14.0 & 29.1 & 11.8 & 22.8 & 5.6  & 16.3 & 1.9 &  14.0 & 2.1 & 11.8 & 1.7 \\
HD 30876  & 38.3 & 18.9 & 34.6 & 13.5 &  30.0 & 5.7 & 21.1 & 5.1 & 19.7 & 3.2 & 16.0 & 3.0 \\
HD 31527  & 37.6 & 20.8 & 30.0 & 10.4 & 21.6 & 3.6 & 17.6 & 2.1  & 14.9 & 2.0 &  11.9 & 2.1 \\
HD 36108  &  12.6 & 3.5 & 11.1 & 1.2 & 9.3 & 0.7 & 8.1 & 0.8  & 7.1 & 1.0 &  5.8 & 1.1 \\
HD 3823   & 8.8 & 2.6 & 7.7 & 2.0 & 6.2 & 1.3 & 5.3 & 1.0 &  4.4 & 0.9 & 3.6 & 1.0 \\
HD 38283  & 7.1 & 0.7 &  6.7 & 0.4 &  5.9 & 0.4 & 5.4 & 0.3 & 4.9 & 0.3 & 4.4 & 0.4 \\
HD 38382  & 14.1 & 4.0 & 12.7 & 2.7 & 9.4 & 1.3 & 7.7 & 1.4 & 5.8 & 1.3 &  4.5 & 1.2 \\
HD 38973  & 10.6 & 1.4  & 9.8 & 0.9 & 8.4 & 0.6 & 7.3 & 0.8 &  6.0 & 0.9 & 4.9 & 0.8 \\
HD 39091  & 7.8 & 1.5 & 7.1 & 0.9 & 6.0 & 0.6 & 5.1 & 0.9 & 4.2 & 0.9 & 3.6 & 0.7 \\
HD 4308   & 5.3 & 1.6 & 4.9 & 1.1 &  4.2 & 0.9 &  3.2 & 0.9 & 2.1 & 0.7 & 1.5 & 0.4 \\
HD 43834  & 9.7 & 3.8 & 9.0 & 3.7 & 6.1 & 1.7 & 4.4 & 2.0 &  3.1 & 1.8 & 2.4 & 1.4 \\
HD 44120  &  8.1 & 0.7 & 7.7 & 0.6 & 6.8 & 0.4 &  6.1 & 0.5 & 5.3 & 0.6 & 4.4 & 0.7 \\
HD 44447  & 16.8 & 3.7 & 15.4 & 2.6 & 13.0 & 1.1 & 11.0 & 0.9 & 10.1 & 0.9 & 8.7 & 1.2 \\
HD 44594  & 13.9 & 5.9 & 11.5 & 1.0 & 10.0 & 0.8 & 8.9 & 1.1 & 6.4 & 2.5 & 5.3 & 2.7 \\
HD 45289  & 11.6 & 1.1 & 10.8 & 0.8 & 9.7 & 0.6 & 9.7 & 2.0 & 7.8 & 0.6 & 6.8 & 0.7 \\
HD 45701  & 19.1 & 5.7 & 17.0 & 3.7 & 14.2 & 1.4 & 11.7 & 1.5 & 10.1 & 1.9 &  8.5 & 2.1 \\
HD 53705  & 5.1 & 1.1 & 4.6 & 0.9 & 3.3 & 0.5 & 3.2 & 0.9 & 2.4 & 0.2 & 2.0 & 0.2 \\
HD 53706  & 6.7 & 0.8 &  6.2 & 0.5 &  5.4 & 0.3 & 4.9 & 0.3 & 4.5 & 0.4 & 4.1 & 0.5 \\
HD 55693  & 16.9 & 1.1  &  15.8 & 1.1 & 13.3 & 0.9 & 11.5 & 1.1 & 9.8 & 1.4 & 8.5 & 1.5 \\
HD 55720  & 17.2 & 9.2 & 17.3 & 9.9 & 17.4 & 3.9 &  13.7 & 2.2 &  11.5 & 2.6 & 10.1 & 2.5 \\
HD 59468  & 9.3 & 0.9 &  8.9 & 0.5 &  7.9 & 0.5 &  7.1 & 0.6 & 6.5 & 0.5 &  6.0 & 0.4 \\
HD 65907A &  9.1 & 1.1 &  8.5 & 1.1 &  6.7 & 0.8 & 4.9 & 1.4 & 3.5 & 1.5 & 2.9 & 1.4 \\
HD 67199  & 14.0 & 1.2 & 13.1 & 1.2 & 11.7 & 1.2 & 10.0 & 1.4 & 7.7 & 2.2 & 6.2 & 2.2 \\
HD 67556  & 58.3 & 35.7 & 58.4 & 34.8 & 64.3 & 24.4 & 49.3 & 9.9  & 38.8 & 7.8 & 29.9 & 7.7 \\
HD 69655  & 31.3 & 20.2  & 27.8 & 13.1 & 25.3 & 8.1 & 20.8 & 5.2 &  17.9 & 4.7 & 16.0 & 4.5 \\
HD 70642  & 7.5 & 0.6 & 7.2 & 0.5 &  6.6 & 0.3 & 6.1 & 0.4 & 5.5 & 0.4 &  4.9 & 0.6 \\
HD 70889  &  44.0 & 6.7 & 42.3 & 11.1 & 34.8 & 3.7  &  26.3 & 5.8 & 14.9 & 5.2 &  9.3 & 2.9 \\
HD 72673  & 11.5 & 10.4 & 9.7 & 7.0 & 7.0 & 3.2 & 5.3 & 1.4 & 4.2 & 1.1 & 3.1 & 0.7 \\
HD 72769  & 19.5 & 10.5 & 18.5 & 7.5 & 14.7 & 2.8 & 11.0 & 3.0 & 10.4 & 1.6 & 8.3 & 1.7 \\
HD 73121  & 11.2 & 0.9 & 10.6 & 0.6 & 9.7 & 0.4 & 9.1 & 0.5 &  8.2 & 0.6 & 7.2 & 0.7 \\
HD 73524  & 10.4 & 2.6 &  8.8 & 1.3 & 6.7 & 0.9 & 5.9 & 1.4 & 4.2 & 1.1 & 3.1 & 1.1 \\
HD 73526  & 13.9 & 6.5 &  11.0 & 2.2 &  8.4 & 1.1 & 7.0 & 0.9 & 5.9 & 1.1 & 4.7 & 1.2 \\
HD 74868  & 13.7 & 1.9 & 12.9 & 1.3  & 11.1 & 1.0 & 9.0 & 1.6 & 7.0 & 2.1 & 5.9 & 2.0 \\
HD 75289  & 10.3 & 0.7 & 9.9 & 0.6 & 9.1 & 0.5 & 8.4 & 0.8 & 7.6 & 0.9 &  6.8 & 1.0 \\
HD 7570   & 10.6 & 1.9 & 9.4 & 0.7  & 7.6 & 0.7 &  6.6 & 0.5 & 5.9 & 0.9 &  5.3 & 1.1 \\
HD 76700  & 18.2 & 6.4 & 16.1 & 5.6 & 12.7 & 2.4 & 10.7 & 1.7 & 8.9 & 1.7 & 7.0 & 1.5 \\
HD 78429  & 23.6 & 8.5 & 19.6 & 3.4 & 15.8 & 1.8 & 13.8 & 1.7  & 11.5 & 2.0 & 9.2 & 1.9 \\
HD 80913  &  38.6 & 8.9 & 36.6 & 7.8 & 31.0 & 5.1 & 26.2 & 3.9 & 22.3 & 4.1 & 18.4 & 4.2 \\
HD 83529A & 24.7 & 10.2 & 22.4 & 7.9 & 19.1 & 3.0 & 15.8 & 1.5 &  13.4 & 1.5 & 11.5 & 1.7 \\
HD 84117  & 7.4 & 0.6 & 6.7 & 0.5 & 5.6 & 0.5 & 5.0 & 0.3 & 4.5 & 0.4 & 3.9 & 0.5 \\
HD 85512  & 19.3 & 11.1 & 16.2 & 5.6 & 14.0 & 3.4 & 11.9 & 2.2 & 10.5 & 1.9 & 9.1 & 1.8 \\
HD 85683  & 41.3 & 31.9 & 42.3 & 29.3 & 44.1 & 15.5 & 31.8 & 5.0 & 25.4 & 4.6 & 20.7 & 4.5 \\
HD 86819  & 23.0 & 3.1 & 20.9 & 1.9 & 17.6 & 1.7 & 14.7 & 2.0 & 12.7 & 1.8 & 10.9 & 1.6 \\
HD 88742  & 30.3 & 4.7 & 28.0 & 2.4 & 25.4 & 2.5  &  20.4 & 3.6  &  17.4 & 4.0 &  14.3 & 4.0 \\
HD 9280   & 50.9 & 28.4 & 50.1 & 24.8 & 33.7 & 4.7  & 24.9 & 4.5 & 22.9 & 2.5  & 19.4 & 3.2  \\
HD 92987  & 7.6 & 0.5 & 7.3 & 0.4 & 6.7 & 0.3 & 6.3 & 0.3 &  5.9 & 0.5 &  5.3 & 0.7 \\
HD 93885  & 14.4 & 1.6 & 13.3 & 0.9 & 11.6 & 0.9 & 10.5 & 0.8 & 9.5 & 0.8 & 8.1 & 0.9 \\
HD 96423  & 11.1 & 1.0 &  10.6 & 0.7 &  9.6 & 0.4 & 8.9 & 0.6 & 8.0 & 0.9 & 7.0 & 1.1 \\
\enddata
\end{deluxetable}

\clearpage


\begin{thebibliography}{}

\bibitem[Butler et al.(2006)]{butler06} Butler, R.~P., Wright, J.~T., 
Marcy, G.~W., et al.\ 2006, \apj, 646, 505

\bibitem[Burke et al.(2015)]{burke15} Burke, C.~J., 
Christiansen, J.~L., Mullally, F., et al.\ 2015, \apj, 809, 8

\bibitem[Carter et al.(2003)]{carter03} Carter, B.~D., Butler, R.~P., 
Tinney, C.~G., et al.\ 2003, \apjl, 593, L43

\bibitem[Chatterjee et al.(2008)]{chat08} Chatterjee, S., Ford, E.~B., 
Matsumura, S., \& Rasio, F.~A.\ 2008, \apj, 686, 580

\bibitem[Chauvin et al.(2012)]{chauvin12} Chauvin, G., Lagrange, A.-M., 
Beust, H., et al.\ 2012, \aap, 542, A41

\bibitem[Cumming et al.(2008)]{cumming08} Cumming, A., Butler, R.~P., 
Marcy, G.~W., Vogt, S.~S., Wright, J.~T., \& Fischer, D.~A.\ 2008, 
\pasp, 120, 531

\bibitem[Cumming \& Dragomir(2010)]{cumming10} Cumming, A., \& 
Dragomir, D.\ 2010, \mnras, 401, 1029

\bibitem[Dressing \& Charbonneau(2015)]{dressing15} Dressing, C.~D., \& 
Charbonneau, D.\ 2015, \apj, 807, 45

\bibitem[Endl et al.(2015)]{endl15} Endl, M., Brugamyer, E.~J., Cochran, 
W.~D., et al.\ 2015, \apj, in press. arXiv:1512.02965

\bibitem[Fischer et al.(2014)]{fischer14} Fischer, D.~A., Marcy, G.~W., 
\& Spronck, J.~F.~P.\ 2014, \apjs, 210, 5

\bibitem[Ford \& Rasio(2008)]{ford08} Ford, E.~B., \& 
Rasio, F.~A.\ 2008, \apj, 686, 621

\bibitem[Fornasier et al.(2007)]{forn07} Fornasier, S., 
Dotto, E., Hainaut, O., et al.\ 2007, Icarus, 190, 622

\bibitem[Fressin et al.(2013)]{fressin13} Fressin, F., Torres, G., 
Charbonneau, D., et al.\ 2013, \apj, 766, 81

\bibitem[Gould et al.(2010)]{gould10} Gould, A., Dong, S., Gaudi, B.~S., 
et al.\ 2010, \apj, 720, 1073

\bibitem[Go{\'z}dziewski \& Migaszewski(2014)]{goz14} 
Go{\'z}dziewski, K., \& Migaszewski, C.\ 2014, \mnras, 440, 3140

\bibitem[Hays et al.(1976)]{hays76} Hays, J.~D., Imbrie, J., \& 
Shackleton, N.~J.\ 1976, Science, 194, 1121

\bibitem[Horner \& Jones(2008)]{hj08} Horner, J., \& 
Jones, B.~W.\ 2008, International Journal of Astrobiology, 7, 251

\bibitem[Horner \& Jones(2009)]{hj09} Horner, J., \& 
Jones, B.~W.\ 2009, International Journal of Astrobiology, 8, 75

\bibitem[Horner \& Jones(2010)]{hj10} Horner, J., \& Jones, B.~W.\ 2010, 
International Journal of Astrobiology, 9, 273

\bibitem[Horner et al.(2012)]{h12} Horner, J., M{\"u}ller, T.~G., \& 
Lykawka, P.~S.\ 2012, \mnras, 423, 2587

\bibitem[Horner \& Jones(2012)]{hj12} Horner, J., \& Jones, B.~W.\ 2012, 
International Journal of Astrobiology, 11, 147

\bibitem[Horner et al.(2015)]{horner15} Horner, J., Gilmore, J.~B., \& 
Waltham, D.\ 2015, arXiv:1511.06043

\bibitem[Howard et al.(2010)]{howard10} Howard, A.~W., Marcy, G.~W., 
Johnson, J.~A., et al.\ 2010, Science, 330, 653

\bibitem[Howard et al.(2012)]{howard12} Howard, A.~W., Marcy, G.~W., 
Bryson, S.~T., et al.\ 2012, \apjs, 201, 15

\bibitem[Jewitt \& Haghighipour(2007)]{jewitt07} Jewitt, D., \& 
Haghighipour, N.\ 2007, \araa, 45, 261

\bibitem[Jones et al.(2010)]{jones10} Jones, H.~R.~A., Butler, R.~P., 
Tinney, C.~G., et al.\ 2010, \mnras, 403, 1703

\bibitem[Kalas et al.(2008)]{kalas08} Kalas, P., Graham, J.~R., Chiang, 
E., et al.\ 2008, Science, 322, 1345

\bibitem[Kopparapu(2013)]{kop13} Kopparapu, R.~K.\ 2013, \apjl, 767, L8

\bibitem[K{\" u}rster et al.(1997)]{kurster97} K{\" u}rster, M.,
Schmitt, J.~H.~M.~M., Cutispoto, G., \& Dennerl, K.\ 1997, \aap, 320,
831

\bibitem[Laakso et al.(2006)]{laakso06} Laakso, T., Rantala, 
J., \& Kaasalainen, M.\ 2006, \aap, 456, 373

\bibitem[Lee et al.(2015)]{lee15} Lee, C.-U., Kim, S.-L., Cha, S.-M., et 
al.\ 2015, IAU General Assembly, 22, 52676

\bibitem[Lykawka \& Horner(2010)]{l10} Lykawka, P.~S., \& Horner, J.\ 
2010, \mnras, 405, 1375

\bibitem[Marois et al.(2008)]{marois08} Marois, C., Macintosh, B., 
Barman, T., et al.\ 2008, Science, 322, 1348

\bibitem[McCarthy et al.(2004)]{mccarthy04} McCarthy, C., 
Butler, R.~P., Tinney, C.~G., Jones, H.~R.~A., Marcy, G.~W., Carter, B., 
Penny, A.~J., \& Fischer, D.~A.\ 2004, \apj, 617, 575

\bibitem[Morbidelli et al.(2005)]{morbi05} Morbidelli, A., Levison, 
H.~F., Tsiganis, K., \& Gomes, R.\ 2005, \nat, 435, 462

\bibitem[Nielsen et al.(2013)]{nielsen13} Nielsen, E.~L., Liu, M.~C., 
Wahhaj, Z., et al.\ 2013, \apj, 776, 4

\bibitem[Owen \& Bar-Nun(1995)]{owen95} Owen, T., \& Bar-Nun, A.\ 1995, 
Icarus, 116, 215

\bibitem[Rameau et al.(2013)]{rameau13} Rameau, J., Chauvin, G., 
Lagrange, A.-M., et al.\ 2013, \apjl, 772, L15

\bibitem[Rowan et al.(2015)]{rowan15} Rowan, D., Meschiari, S., 
Laughlin, G., et al.\ 2015, \apj, in press. arXiv:1512.00417

\bibitem[Sheppard \& Jewitt(2003)]{sheppard03} Sheppard, S.~S., \& 
Jewitt, D.~C.\ 2003, \nat, 423, 261

\bibitem[Vigan et al.(2015)]{vigan15} Vigan, A., Bonnefoy, M., Ginski, 
C., et al.\ 2015, arXiv:1511.04076

\bibitem[Vinogradova \& Chernetenko(2015)]{v15} Vinogradova, T.~A., \& 
Chernetenko, Y.~A.\ 2015, Solar System Research, 49, 391


\bibitem[Wittenmyer et al.(2010)]{foreverpaper} Wittenmyer, R.~A., 
O'Toole, S.~J., Jones, H.~R.~A., et al.\ 2010, \apj, 722, 1854

\bibitem[Wittenmyer et al.(2011a)]{jupiters} Wittenmyer, R.~A., Tinney, 
C.~G., O'Toole, S.~J., Jones, H.~R.~A., Butler, R.~P., Carter, B.~D., \& 
Bailey, J.\ 2011a, \apj, 727, 102

\bibitem[Wittenmyer et al.(2011b)]{etaearth} Wittenmyer, R.~A., 
Tinney, C.~G., Butler, R.~P., et al.\ 2011b, \apj, 738, 81 

\bibitem[Wittenmyer et al.(2012)]{142paper} Wittenmyer, R.~A., Horner, 
J., Tuomi, M., et al.\ 2012, \apj, 753, 169

\bibitem[Wittenmyer et al.(2013a)]{witt13} Wittenmyer, R.~A., Tinney, 
C.~G., Horner, J., et al.\ 2013a, \pasp, 125, 351

\bibitem[Wittenmyer et al.(2013b)]{songhu} Wittenmyer, R.~A., 
Wang, S., Horner, J., et al.\ 2013b, \apjs, 208, 2 

\bibitem[Wittenmyer et al.(2014a)]{2jupiters} Wittenmyer, R.~A., Horner, 
J., Tinney, C.~G., et al.\ 2014a, \apj, 783, 103

\bibitem[Wittenmyer et al.(2014b)]{gj832} Wittenmyer, R.~A., 
Tuomi, M., Butler, R.~P., et al.\ 2014b, \apj, 791, 114 

\bibitem[Wittenmyer \& Marshall(2015)]{debris} Wittenmyer, 
R.~A., \& Marshall, J.~P.\ 2015, \aj, 149, 86

\bibitem[Wright et al.(2011)]{wright11} Wright, J.~T., et al.\ 2011, 
\pasp, 123, 412

\bibitem[Zechmeister \& K{\" u}rster(2009)]{zk09} Zechmeister, M., K{\" 
u}rster, M.\ 2009, \aap, 496, 577

\bibitem[Zurlo et al.(2015)]{zurlo15} Zurlo, A., Vigan, A., Galicher, 
R., et al.\ 2015, arXiv:1511.04083



\end{thebibliography}
\end{document}